\documentclass[usenatbib]{mn2e}

\title[Central Black Holes, Bar Dynamics, and Dark Matter Halos]{On the Link Between Central Black Holes, Bar Dynamics, and Dark Matter Halos in Spiral Galaxies}
\author[P. Treuthardt et al.]{Patrick Treuthardt$^{1,2}$\thanks{E-mail:
pxtreuthardt@ualr.edu}, Marc S. Seigar$^{1,2}$, Amber D. Sierra$^1$, Ismaeel Al-Baidhany$^1$,
\newauthor Heikki Salo$^3$, Daniel Kennefick$^{2,4}$, Julia Kennefick$^{2,4}$, and Claud H.~S. Lacy$^{2,4}$\\
$^1$Department of Physics \& Astronomy, University of Arkansas at Little
Rock, 2801 S.\ University Avenue, Little Rock, AR 72204, U.S.A.\\
$^2$Arkansas Center for Space and Planetary Sciences, 202 Old Museum
Building, University of Arkansas, Fayetteville, AR 72701, U.S.A.\\
$^3$Division of Astronomy, Department of Physical Sciences, University
of Oulu, Oulu, FIN-90014, Finland\\
$^4$Department of Physics, University of Arkansas, 835 West Dickson
Street, Fayetteville, AR 72701, U.S.A.}

\begin{document}

\date{Submitted 2011 June 24}

\pagerange{\pageref{firstpage}--\pageref{lastpage}} \pubyear{2002}

\maketitle

\label{firstpage}

\begin{abstract}
The discovery of a relationship between 
supermassive black hole (SMBH) mass and spiral arm pitch angle 
($P$) is evidence that SMBHs are tied to the overall secular 
evolution of a galaxy. The discovery of SMBHs in late-type galaxies
with little or no bulge suggests that an underlying correlation between
the dark matter halo concentration and SMBH mass ($M_{BH}$) exists, rather than between
the bulge mass and $M_{BH}$. In this paper we measure $P$ using a two-dimensional fast fourier transform and estimate the bar pattern speeds of 40 barred spiral 
galaxies from the Carnegie-Irvine Galaxy Survey. The pattern speeds were derived by estimating the
gravitational potentials of our galaxies from $K_s$-band images and using them to produce dynamical simulation models.
The pattern speeds allow us to identify those galaxies with low central dark halo densities, or fast rotating bars, while $P$ provides an estimate
of $M_{BH}$.   We find that a wide range of $M_{BH}$ exists in 
galaxies with low central dark matter halo densities, which appears to support other theoretical results. 
We also find that galaxies 
with low central dark halo densities appear to follow more predictable trends 
in $P$ versus de Vaucouleurs morphological type ($T$) and bar strength versus $T$ than barred galaxies in 
general. The empirical relationship between $M_{BH}$ and total gravitational mass of a galaxy ($M_{tot}$) allows us to predict the minimum $M_{tot}$ 
that will be observationally measured of our fast bar galaxies. These predictions will be investigated in a subsequent paper.
\end{abstract}


\section{Introduction}

The connection between the overall morphology and dynamics of galaxies and the
centrally located supermassive black holes (SMBHs) they harbor provides a 
fundamental constraint in the study of galaxy formation and evolution.
One well-known example of this link is seen in the correlation 
between SMBH mass ($M_{BH}$) and the large-scale stellar velocity dispersion 
($\sigma$) of the bulge/spheroid encompassing it (Gebhardt et al. 2000; 
Ferrarese \& Merritt 2000).  Likewise, the discovery of a relationship between 
$M_{BH}$ and spiral arm pitch angle 
($P$; Seigar et al. 2008) suggests that SMBHs are tied to the overall secular 
evolution of a galaxy. Seigar et al.\
(2005, 2006) have already shown that there is a strong correlation between 
$P$ and rotation curve shear rate ($S$), where small values of $P$ correspond 
to large $S$ (see their Figure 3) and a large central mass concentration. More
recent studies have further indicated that SMBHs are linked to secular processes 
(e.g. Cisternas et al. 2011, Orban de Xivry et al. 2011).

The stellar velocity dispersion of the bulge/spheroid is commonly linked to $M_{BH}$ (Gebhardt et al. 2000; Ferrarese \& Merritt 2000).
For example, M33 is a bulgeless late-type galaxy with the lowest nuclear BH mass ever measured with an upper mass limit of 1500 
$M_{\sun}$ (Gebhardt et al. 2001). The discovery of SMBHs in the 
centers of late-type galaxies with little or no bulge (Satyapal et al. 2007, 2008) 
suggests that the size, or mass, of SMBHs may be tied to the dark matter
halo virial mass, or concentration (see, e.g., van den Bosch et al. 2007 for 
a relation). This hints at the cause of the $M_{BH}-P$ connection and the reason why M33 fits 
in this relation. Comparisons of the corrected central velocity dispersion with the 
maximum rotation velocity of disk galaxies also reveal a good correlation 
between $M_{BH}$ and the mass of the dark matter halo (Ferrarese 2002).  
Additionally, bulge size and dark halo mass show a direct correlation in general, but 
the relationship exhibits a lot of scatter (see, e.g., Ho 2007). Recently, 
Booth \& Schaye (2010) have shown through simulations that $M_{BH}$ appears 
to be linked to halo binding energy rather than the halo mass, with the 
central halo concentration playing a role in the relationship's scatter. Other 
recent papers have also argued both for (Seigar 2011; Volonteri et 
al.\ 2011) and against (Kormendy \& Bender 2011) a coupling of SMBH masses and 
the properties of the dark matter halo. 

The secular evolution of a barred spiral galaxy is chiefly influenced by the 
rotation rate of the non-axisymmetric component (e.g., 
Lynden-Bell \& Kalnajs 1972; Kalnajs 1991; Kormendy \& Kennicutt 2004).  
Simulation models have shown that after a bar initially forms, the pattern 
speed ($\Omega_{p}$) remains fast when embedded in a dark matter halo with a
low central density (Debattista \& Sellwood 2000).  This is due to reduced 
dynamical friction between the two components that would otherwise cause a 
rapid decrease in $\Omega_p$.  $\Omega_{p}$ also determines the locations of 
important resonance regions in galaxies with a single perturbation mode, such
as a bar or grand design spiral.  This allows the rotation of a bar to be 
described by the ratio $\mathcal{R} = R_{CR}/R_{bar}$, where $R_{CR}$ is the 
corotation, or 1:1, resonance (CR) radius and $R_{bar}$ is the bar 
semimajor axis length. When $1.0 \leq \mathcal{R} \leq 1.4$ a
bar is deemed to be a fast rotator, while $\mathcal{R} > 1.4$ implies a 
slow rotating bar. 
Theoretical arguments by Contopoulos (1980) state that self-consistent bars 
must have $\mathcal{R} > 1.0$.

About 50\% of all spiral galaxies display either a full, broken, or partial 
ring 
shaped pattern in their light distribution (Buta \& Combes 1996). Although 
rings are observed in some non-barred galaxies, they are typically associated 
with barred galaxies due to the ease at which they form through bar-driven 
gravity torques (see Grouchy et al. 2010 and references therein). Theoretical 
arguments suggest that inner rings in barred spiral galaxies are 
features of the inner ultraharmonic, or 4:1, resonance (IUHR; Schwarz 1984; 
Byrd et al. 1998). Athanassoula et al. (1982) have discussed that for a flat 
rotation curve, where $\kappa = \sqrt{2}\Omega$, the ratio of the outer
Lindblad, or 2:1, resonance (OLR) to IUHR 
radius is 
\begin{equation}
R_{OLR}/R_{IUHR}= \frac{1+\sqrt{2}/2}{1-\sqrt{2}/4}=2.64,
\end{equation}
while the ratio of the OLR to CR radius is 
\begin{equation}
R_{OLR}/R_{CR}=1+\sqrt{2}/2=1.71.
\end{equation} 
This implies that the ratio of the CR to IUHR radius is 
\begin{equation}
R_{CR}/R_{IUHR}=\frac{1}{1-\sqrt{2}/4}=1.55.
\end{equation}
Since bars are encircled by inner rings, 
if present, $R_{bar}$ must be similar to the inner ring radius. This suggests 
that $\mathcal{R} \sim 1.6$ or that bars are slow rotators in general.  
Statistics of observational data show that the median ratio of outer to inner 
ring diameter is 2.2 (Buta 1995) or the approximate mean of the theoretical 
OLR:IUHR and OLR:CR values. If outer rings in general can be linked to the
OLR, this suggests a nearly even distribution of galaxies with fast or slow 
bars. Recent evidence suggests that outer rings may not always 
be linked to the OLR as expected though (Treuthardt et al. 2008).   

$\Omega_{p}$, or $\mathcal{R}$, can be estimated using a number of methods
(e.g., Canzian 1993; Buta \& Combes 1996; Puerari 
\& Dottori 1997; Salo et al. 1999; Weiner et al. 2001; Egusa et al. 2004; 
Zhang \& Buta 2007) with the most direct being that of Tremaine \& Weinberg
(1984; TW).  Much emphasis 
has been placed on this method recently but it has been suggested that 
morphological matching through
dynamical simulations yield similar estimates while lacking the limited 
applicabilty of the TW method (Treuthardt et al. 2009). 

The goal of our study is to estimate the bar pattern speeds and measure the 
spiral arm pitch angles of 40 barred spiral galaxies.  This will indicate 
which of these galaxies have low central dark halo densities and provide an 
estimate 
of the central SMBH mass, respectively. Based on the simulation 
results of Booth \& Schaye (2010), we expect to find that for a given halo mass, SMBHs in galaxies with fast bars will have a 
less-than-average mass due to anemic growth within the low central density dark halo. 
This means that the total luminous and dark mass of these galaxies 
($M_{tot}$) predicted from the $M_{BH}-M_{tot}$ relationship 
(Bandara et al. 2009) would be the minimum expected observationally measured value.
We will investigate this claim with measurements of $M_{tot}$ in a forthcoming 
paper. 


\section{Analysis}

\subsection{Pitch Angle Measurements}

$B$-band images from the Carnegie-Irvine Galaxy Survey (CGS; Ho et al. 2011) 
were used to determine $P$ for our 
sample of barred spiral galaxies, since $P$ has been shown to be independent
of the wavelength at which it is measured (Seigar et al.\ 2006).
These deep, high resolution images highlight 
the young stellar component of the spiral arms and offer a larger field of 
view than the $K_{s}$-band images, thereby revealing more of the outer 
spiral structure of each galaxy. 
The orientation and ellipticity values of the outer 
$B$-band isophotes were determined with ELLIPSE in IRAF and used to derive an
inclination (Hubble 1926; see Table 1).  The images were then deprojected to 
face-on by assuming that spiral galaxy disks are intrinsically circular.
Additionally, the output from ELLIPSE was used to estimate $R_{bar}$ in each 
galaxy.  The radius, near the visible ends of 
the bar, where the ellipticity reaches a maximum was selected as $R_{bar}$. 

\begin{table*}
 \centering
 \begin{minipage}{140mm}
        \begin{tabular}{@{}lccccccccc@{}}
  \hline
   Galaxy & $i$ & $R_{bar}$ & $D_{25}$ & $\mathcal{R}$ & $P$ & $M_{BH}$ & $M_{tot}$ & $T$ & $Q_{g}$\\
   (1) & (2) & (3) & (4) & (5) & (6) & (7) & (8) & (9) & (10)\\
  \hline
ESO 121-026 & 49.8 & 20.2 (3.13) & 22.3 & 1.38 $\pm$ 0.05 & 10.5 $\pm$ 1.2$^{a}$ & 343 $\pm$ 119 & 38.3 $\pm$ 8.0 & 3.9 $\pm$ 0.4 & 0.258\\
ESO 380-001 & 60.7 & 65.5 (1.29) & 25.3 & 1.48 $\pm$ 0.06 & 32.3 $\pm$ 2.0       & 12.6 $\pm$ 2.8 & 4.55 $\pm$ 1.39 & 2.8 $\pm$ 0.7 & 0.318\\
ESO 506-004 & 62.9 & 15.0 (4.28) & 37.4 & 1.74 $\pm$ 0.08 & 14.0 $\pm$ 1.0       & 148 $\pm$ 31 & 22.3 $\pm$ 0.3 & 2.5 $\pm$ 1.0 & 0.131\\
IC 1953     & 48.2 & 29.3 (3.37) & 15.8 & 1.83 $\pm$ 0.06 & 33.0 $\pm$ 1.6       & 11.7 $\pm$ 2.2 & 4.33 $\pm$ 1.46 & 6.1 $\pm$ 0.9 & 0.380\\
IC 2367     & 40.9 & 15.5 (2.77) & 26.4 & 2.24 $\pm$ 0.06 & 15.8 $\pm$ 0.7       & 104 $\pm$ 14 & 17.8 $\pm$ 1.8 & 3.1 $\pm$ 0.7 & 0.265\\
IC 2560     & 61.7 & 39.9 (8.50) & 45.3 & 1.47 $\pm$ 0.06 & 28.4 $\pm$ 4.2       & 18.7 $\pm$ 8.9 & 5.88 $\pm$ 0.54 & 3.4 $\pm$ 0.6 & 0.112\\
IC 5240     & 48.3 & 39.6 (4.04) & 19.4 & 1.50 $\pm$ 0.06 & 24.4 $\pm$ 0.9       & 29.2 $\pm$ 3.2 & 7.83 $\pm$ 2.15 & 1.0 $\pm$ 0.3 & 0.193\\
IC 5273     & 48.3 & 23.3 (1.58) & 12.6 & 1.59 $\pm$ 0.07 & 33.5 $\pm$ 2.8       & 11.0 $\pm$ 4.1 & 4.17 $\pm$ 0.91 & 5.7 $\pm$ 0.9 & 0.258\\
NGC 0151    & 63.7 & 18.1 (4.05) & 49.9 & 1.40 $\pm$ 0.06 & 36.1 $\pm$ 1.5$^{a}$ & 6.98 $\pm$ 2.71 & 3.11 $\pm$ 0.81 & 4.0 $\pm$ 0.5 & 0.105\\
NGC 0337    & 41.2 & 14.8 (1.30) & 15.6 & 1.27 $\pm$ 0.07 & 32.9 $\pm$ 1.4       & 11.8 $\pm$ 1.9 & 4.36 $\pm$ 1.54$^*$ & 6.7 $\pm$ 0.8 & 0.800\\
NGC 0782    & 30.6 & 22.8 (8.87) & 57.3 & 1.36 $\pm$ 0.06 & 23.7 $\pm$ 1.9       & 31.8 $\pm$ 7.6 & 8.27 $\pm$ 1.49 & 3.0 $\pm$ 0.3 & 0.126\\
NGC 0945    & 41.0 & 17.6 (4.89) & 34.1 & 1.66 $\pm$ 0.06 & 15.5 $\pm$ 1.3       & 110 $\pm$ 28 & 18.4 $\pm$ 0.4 & 5.0 $\pm$ 0.5 & 0.301\\
NGC 1022    & 20.5 & 18.4 (1.49) & 12.5 & 1.87 $\pm$ 0.06 & 13.7 $\pm$ 3.9       & 158 $\pm$ 172 & 23.2 $\pm$ 17.9 & 1.1 $\pm$ 0.3 & 0.122\\
NGC 1317    & 28.3 &  6.5 (0.787) & 22.4 & 1.84 $\pm$ 0.06 & 10.0 $\pm$ 0.6       & 395 $\pm$ 70 & 42.0 $\pm$ 4.6 & 0.8 $\pm$ 0.5 & 0.040\\
NGC 1723    & 39.4 & 18.9 (4.61) & 44.2 & 1.87 $\pm$ 0.06 & 21.3 $\pm$ 3.4       & 43.4 $\pm$ 22.0 & 10.1 $\pm$ 0.8 & 1.2 $\pm$ 0.6 & 0.115\\
NGC 1832    & 41.0 & 12.7 (1.64) & 19.0 & 1.61 $\pm$ 0.08 & 25.3 $\pm$ 6.1       & 26.3 $\pm$ 22.5 & 7.31 $\pm$ 2.10 & 4.0 $\pm$ 0.1 & 0.226\\
NGC 2223    & 24.2 & 18.6 (3.48) & 31.6 & 1.58 $\pm$ 0.06 & 22.8 $\pm$ 3.5       & 35.6 $\pm$ 17.2 & 8.90 $\pm$ 0.73 & 3.8 $\pm$ 0.9 & 0.095\\
NGC 2525    & 44.0 & 14.2 (1.72) & 22.4 & 1.78 $\pm$ 0.06 & 22.9 $\pm$ 1.9       & 35.1 $\pm$ 8.7 & 8.82 $\pm$ 1.43 & 5.2 $\pm$ 0.5 & 0.231\\
NGC 2763    & 36.5 &  4.1 (0.595) & 21.4 & 1.67 $\pm$ 0.08 & 20.5 $\pm$ 2.5       & 48.6 $\pm$ 18.1 & 10.9 $\pm$ 0.6 & 5.7 $\pm$ 0.7 & 0.235\\
NGC 3124    & 35.5 & 15.0 (3.84) & 41.3 & 1.33 $\pm$ 0.05 & 20.5 $\pm$ 1.9       & 48.6 $\pm$ 13.5 & 10.9 $\pm$ 1.1$^*$ & 3.9 $\pm$ 0.5 & 0.209\\
NGC 3275    & 28.2 & 18.6 (4.24) & 38.6 & 1.43 $\pm$ 0.06 & 30.0$^{b}$           & 15.9 & 5.29 $\pm$ 2.21 & 1.9 $\pm$ 0.5 & 0.108\\
NGC 3347    & 33.7 & 13.0 (2.83) & 53.3 & 1.73 $\pm$ 0.05 & 37.8 $\pm$ 4.1       & 3.63 $\pm$ 5.35 & 2.04 $\pm$ 1.06 & 3.5 $\pm$ 0.8 & 0.116\\
NGC 3450    & 20.2 & 25.1 (7.18) & 45.1 & 1.65 $\pm$ 0.06 & 11.8 $\pm$ 0.4$^{a}$ & 244 $\pm$ 24 & 30.8 $\pm$ 0.3 & 3.1 $\pm$ 0.4 & 0.152\\
NGC 3513    & 40.5 & 24.1 (2.46) & 17.2 & 1.24 $\pm$ 0.07 & 26.7 $\pm$ 3.6$^{a}$ & 22.4 $\pm$ 9.4 & 6.61 $\pm$ 0.66$^*$ & 5.1 $\pm$ 0.4 & 0.540\\
NGC 3660    & 42.4 & 16.3 (4.32) & 40.9 & 1.75 $\pm$ 0.10 & 20.4 $\pm$ 2.3       & 49.3 $\pm$ 16.9 & 11.0 $\pm$ 0.6 & 3.9 $\pm$ 0.5 & 0.145\\
NGC 3887    & 49.1 & 16.8 (1.75) & 20.2 & 1.89 $\pm$ 0.08 & 24.7 $\pm$ 3.7$^{a}$ & 28.2 $\pm$ 13.3 & 7.65 $\pm$ 0.64 & 3.9 $\pm$ 0.5 & 0.143\\
NGC 4050    & 50.0 & 30.0 (4.20) & 28.5 & 1.65 $\pm$ 0.06 &  8.9 $\pm$ 1.2$^{a}$ & 555 $\pm$ 232 & 52.4 $\pm$ 17.2 & 2.1 $\pm$ 0.6 & 0.203\\
NGC 4593    & 45.3 & 57.0 (11.4) & 28.8 & 1.23 $\pm$ 0.06 & 22.9 $\pm$ 3.8       & 35.1 $\pm$ 18.6 & 8.82 $\pm$ 0.82$^*$ & 3.0 $\pm$ 0.4 & 0.181\\
NGC 5135    & 28.4 & 33.2 (9.53) & 41.3 & 1.27 $\pm$ 0.06 & 15.8 $\pm$ 1.5       & 104 $\pm$ 30 & 17.8 $\pm$ 0.5$^*$ & 2.4 $\pm$ 0.6 & 0.099\\
NGC 5156    & 17.1 &  9.3 (1.96) & 31.8 & 1.61 $\pm$ 0.11 & 20.9 $\pm$ 1.6       & 45.9 $\pm$ 10.5 & 10.5 $\pm$ 1.5 & 3.5 $\pm$ 0.9 & 0.239\\
NGC 5339    & 40.0 & 26.9 (5.33) & 21.6 & 1.75 $\pm$ 0.08 & 16.2 $\pm$ 2.0       & 96.6 $\pm$ 36.6 & 16.9 $\pm$ 0.9 & 1.0 $\pm$ 0.5 & 0.255\\
NGC 5728    & 53.3 & 43.8 (8.72) & 37.8 & 1.56 $\pm$ 0.06 & 10.3 $\pm$ 2.3       & 362 $\pm$ 278 & 39.8 $\pm$ 21.4 & 1.2 $\pm$ 0.7 & 0.232\\
NGC 5938    & 28.3 & 23.1 (5.47) & 32.6 & 1.64 $\pm$ 0.06 & 36.1 $\pm$ 9.3       & 7.0 $\pm$ 11.1 & 3.11 $\pm$ 1.73 & 4.3 $\pm$ 0.6 & 0.151\\
NGC 6782    & 23.1 & 26.7 (6.76) & 37.3 & 1.20 $\pm$ 0.06 & 13.4 $\pm$ 1.2       & 168 $\pm$ 45 & 24.2 $\pm$ 1.3$^*$ & 1.0 $\pm$ 0.9 & 0.060\\
NGC 6923    & 50.8 & 11.7 (2.00) & 25.8 & 1.31 $\pm$ 0.06 & 23.4 $\pm$ 3.7       & 33.0 $\pm$ 16.5 & 8.47 $\pm$ 0.74$^*$ & 3.1 $\pm$ 0.6 & 0.076\\
NGC 7059    & 71.1 & 41.7 (4.42) & 23.6 & 1.72 $\pm$ 0.08 & 17.5 $\pm$ 1.3       & 77.1 $\pm$ 17.0 & 14.6 $\pm$ 1.2 & 5.7 $\pm$ 1.0 & 0.210\\
NGC 7070    & 35.4 & 10.4 (1.49) & 21.6 & 1.67 $\pm$ 0.16 & 30.2 $\pm$ 5.0       & 15.6 $\pm$ 9.0 & 5.22 $\pm$ 0.67 & 6.0 $\pm$ 0.3 & 0.183\\
NGC 7218    & 66.5 &  8.8 (0.774) & 13.6 & 1.52 $\pm$ 0.06 & 34.3 $\pm$ 10.4      & 9.9 $\pm$ 15.5 & 3.89 $\pm$ 2.27 & 5.6 $\pm$ 1.2 & 0.254\\
NGC 7392    & 55.7 & 19.2 (3.61) & 25.3 & 1.44 $\pm$ 0.06 & 24.6 $\pm$ 2.0$^{a}$ & 28.5 $\pm$ 6.9 & 7.71 $\pm$ 1.47 & 3.8 $\pm$ 0.7 & 0.177\\
NGC 7723    & 48.5 & 18.9 (1.91) & 25.2 & 1.44 $\pm$ 0.06 & 15.7 $\pm$ 2.3$^{a}$ & 106 $\pm$ 49 & 18.0 $\pm$ 2.2 & 3.1 $\pm$ 0.5 & 0.243\\
\hline
\end{tabular}
\caption{Estimated Galaxy Parameters.}{Explanation of columns: (1) Galaxy designation; (2) inclination in degrees derived from fitting ellipses to $B$-band isophotes; (3) bar semimajor axis length in the sky plane, estimated from ellipse fitting, and measured in arcseconds and converted to kpc (paranthetical values) using the default (i.e. $H_0 = 73.0$, $\Omega_{matter} = 0.27$, and $\Omega_{vacuum} = 0.73$) cosmology-corrected scaling factors from NED; (4) $D_{25}$ average isophotal diameters at 25 mag/arcsec$^{2}$ in the $B$-band taken from HyperLeda (Paturel et al. 2003) and converted to kpc as in (3); (5) estimated value and 1$\sigma$ error of $R_{CR}/R_{bar}$; (6) spiral arm pitch angle measured by the authors of this paper, $^{a}$Seigar et al. 2006, or by $^{b}$Block et al. 1999; (7) SMBH mass and mean error in $10^5$ solar masses derived from $P$ using equation 2 of Seigar et al. 2008; (8) total mass of the galaxy and mean error in $10^{11}$ solar masses derived from $M_{BH}$ using equation 8 of Bandara et al. 2009 ($^*$indicates an estimated minimum halo mass due to $\mathcal{R} \leq 1.4$ according to Booth \& Schaye 2010); (9) de Vaucouleur morphological type index from HyperLeda (Paturel et al. 2003); (10) the maximum of $Q_{T}$ in the bar region estimated from mass models derived from $K_{s}$-band images.}
\end{minipage}
\end{table*}

$P$ was measured using two-dimensional fast Fourier decompositions of the 
deprojected $B$-band images and assuming logarithmic spirals 
(Schr\"{o}der et al. 1994; see Table 1). The 
Fourier fits were applied to visually selected annulus regions that range from 
just beyond the ends of the 
bars to the outer limits of the visible arms. $P$ is then determined from peaks
in the Fourier spectra as this is the most powerful method for finding 
periodicity in a distribution (Consid\`{e}re \& Athanassoula 1988; Garcia-Gomez 
\& Athanassoula 1993). The largest source of error 
in measuring $P$ presumably comes from this selection of radial range 
(Seigar et al.\ 2005, 2008), but galaxies with large inclinations 
($>$ 60$^{\circ}$) also have variances in $P$ of 10\% (Block et al. 1999).
For details on how accurate our method is for measuring $P$, see Davis et al. (2012).

\subsection{Determining the Potential}

Attempting to recreate the observed morphologies of our sample of galaxies 
through simulation modeling requires us to first derive gravitiatonal 
potentials. 
Assuming that the near-IR light distribution follows the mass distribution, 
we applied a mass-to-light ratio ($M/L$) correction to the CGS 
near-IR $K_{s}$-band images of our galaxies to obtain surface mass 
densities.  The $M/L$ correction was estimated from the azimuthally averaged
CGS $B-V$ color index of each galaxy and applying the $K$-band correction 
given by Bell et al.\ (2003).

Next, a two-dimensional iterative bulge-bar-disk decomposition 
program
(see Laurikainen et al.\ 2004 for a description) was applied in 
order to more accurately model the bulge component of the galaxies.  The bulge
component was then removed, each galaxy was deprojected to a face-on 
orientation, and the bulge was then added back in.  We assume the bulges to
be spherical, so this process eliminates 
any bulge stretching from the deprojection process.

The deprojected image was then approximated by the Fourier decomposition method
described by Laurikainen \& Salo (2002) and the disk gravity was derived from 
the even $m = 0$ to $8$ components.  The empirically derived ratio of vertical 
to radial scalelength corresponding to morphological type (de Grijs 1998) was 
used to apply a constant scaleheight throughout the disk.  Our mass models also 
include a dark halo component based on the universal rotation curve of Persic
et al. (1996).  The halo profile was calculated using $L/L_{*}$ 
derived from HyperLeda (Paturel et al.\ 2003) data and the average distances 
calculated from NED (NASA/IPAC Extragalactic Database) data.

A limited amount of kinematic data exists for our sample of galaxies 
(Mathewson et al. 1992; M\'{a}rquez et al. 2004; Catinella et al. 2005).  What data we have
are chiefly in the form of H$\alpha$ rotation curves which only extend to 
about 80\% of the visible galaxy disk (Mathewson et al. 1992) and do not extend 
into the dark matter dominated regions of our mass models. The available data, 
therefore, does little to constrain the outer mass-modeled 
circular speed curves. The upper plot of Figure 1 shows an example of 
azimuthally averaged 
circular speed curve derived from the mass model of ESO 506-004.  The mass 
model was scaled so that the circular speed curve would more accurately fit 
the rotation curve data derived from Figure 3 of Mathewson et al. (1992). The
lower plot in Figure 1 shows the Lindblad precession frequency curves derived
from the circular speed curve.  Our estimate of $\Omega_{p}$, average 
error, and bar semimajor axis radius for this galaxy
are overlayed.  The radial regions where specific resonances in the linear 
(epicyclic) approximation are located are 
indicated where the horizontal line corresponding to $\Omega_{p}$ intersects 
each frequency curve.  For this galaxy, we estimate that the bar extends to 
the IUHR and not to the CR.

\begin{figure*}
\includegraphics{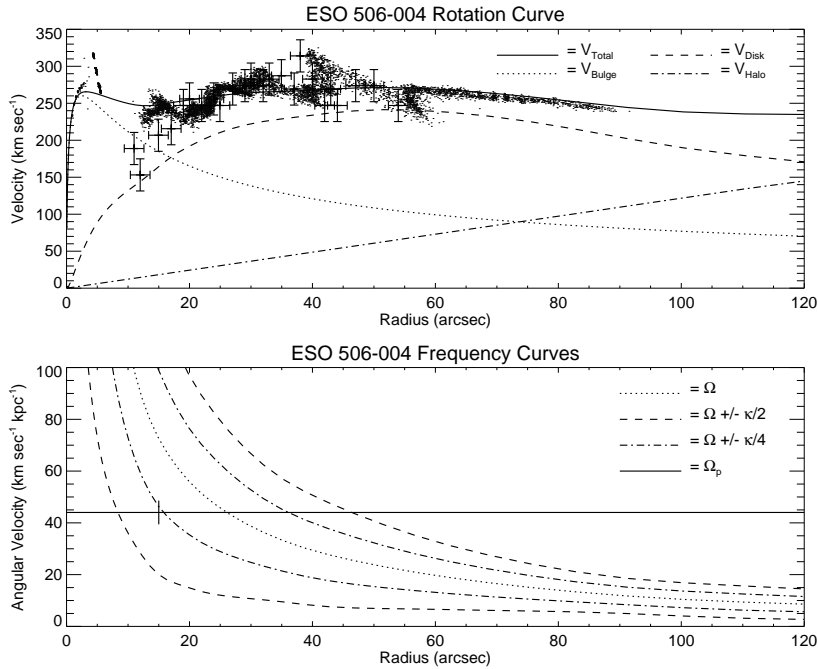}
\vspace*{8.5cm}
\caption{Plots of the rotation curve (top) and Lindblad precession frequency 
curves (bottom) for ESO 506-004. Using the same scaling as in column 3 of Table 1, 
10 arcseconds corresponds to 2.85 kpc. The upper plot shows the total, bulge, disk, 
and dark matter halo components of the azimuthally averaged mass modeled 
circular speed. The dark matter halo component was derived from the universal
rotation curve model of Persic et al. (1996). H$\alpha$ rotation curve points 
derived from Fig. 3 of Mathewson et al. (1992) are overlayed with the estimated 
errors in their values. The inner region of this specific galaxy appears to 
have little hydrogen gas. The modeled major axis velocity profile (points), 
derived from the modeled gas particle distribution of our best simulation, is 
also overlayed. The velocity profile data were extracted from within $3^{\circ}$
of the major axis position angle and corrected for inclination.  The modeled 
velocity profile appears to agree with the observed rotation curve.
The lower plot shows $\Omega$ (the circular frequency), $\Omega \pm 
\kappa/2$ (where $\kappa$ is the epicyclic frequency), $\Omega \pm \kappa/4$, 
and the bar pattern speed ($\Omega_{p} = 44.0$ km s$^{-1}$ 
kpc$^{-1}$) corresponding to our estimate of $\mathcal{R} = 1.74$. The 
position of the short vertical line associated with 
$\Omega_{p}$ indicates the estimated bar semimajor axis length,
while the length of the line
corresponds to the average error in $\Omega_{p}$. For this galaxy, it appears
the bar semimajor axis extends to the inner 4:1 resonance.}
\label{landfig}
\end{figure*}

Once gravitational potentials were determined for our sample, we were
also able to estimate the strength of the bar in each galaxy.  This was done
by calculating two dimensional maps of the radial ($F_{R}$) and tangential 
($F_{T}$) forces. The tangential forces induced by the bar were calculated
using a Polar method discussed by Laurikainen \& Salo (2002) and Laurikainen
et al. (2004).  The Polar method was applied to an azimuthal Fourier 
decomposition of the intensity and the even components, up to m = 20, were then
converted to the corresponding potential components (Salo et al. 1999). The 
ratio of the maximum of $F_{T}$ to the 
azimuthally averaged $F_{R}$ at each radius produces a radial profile of the 
distance dependent maximum tangential force.  The maximum of the this profile 
gives a single measure of the bar strength, $Q_{g}$.  $Q_{g}$, however, can be
affected by spiral arm torques, making it a convolution of the 
bar ($Q_{B}$) and spiral strength ($Q_{S}$; Buta et al. 2005). Therefore, 
$Q_{g}$ can be thought of, at worst, as an upper limit to the 
true bar strength, $Q_{B}$.

\subsection{Simulations}

Our sample of galaxies was modeled by simulating the behaviour of 
a two-dimensional disk of inelastically colliding (or ``sticky''), non-self-gravitating gas particles in our 
predetermined potentials. The details of the simulation code we used can be 
found in Salo et al. (1999) and Salo (1991). The advantage of using a sticky particle code over
a smoothed particle hydrodynamic (SPH) code, for example, is that it better approximates the cold gas component of a 
galaxy (Merlin \& Chiosi 2007). In SPH, gas particles are subject to nongravitational forces, such as pressure gradients,
making it better suited for modeling the warm gas phase. Therefore, it is appropriate to use sticky particle models for 
matching observed morphological components containing star formation, such as resonance rings and spiral arms. The 
overall modeling process closely followed that of Treuthardt 
et al. (2009). For example, we produced snapshots of 100,000 gas particles at 
the bar rotation period of interest by simulating the behaviour of a two-dimensional disk of 20,000 gas particles and 
aligning and coadding individual snapshots within $\pm$ 0.2 bar rotation periods in 0.1 period increments. We also assumed 
that each galaxy has only one pattern speed, that 
of the bar, and $\Omega_{p}$ was the main parameter that was varied.  
With an estimate of $R_{bar}$ and using the Lindblad precession frequency 
curves derived from our potentials, we are able to link $\Omega_{p}$ to
 $\mathcal{R}$ via $R_{CR}$.  In doing so, we varied $\Omega_{p}$ by allowing $\mathcal{R}$ 
to increase from 1.0 by increments of 0.1. Treuthardt et al. (2008) investigated
the morphological effects of differing the gravitational potential used in the simulations of NGC 1433.  They found
that the different M/L models of Bell \& de Jong (2001) had little effect on the modeled morphology of the galaxy.
The amplitude of the bar potential was also incrementally altered in order to account for possible uncertainties in the 
bar height or halo contribution. The morphology of the resultant models were markedly different from each other in terms of
inner ring shape and outer spiral morphology (see Figure 9 of Treuthardt et al. 2008). This implies that the use of 
grossly inaccurate gravitational potentials will not produce models with morphologies that can be easily matched to 
observations. Further examples of morphological variation within sticky particle models of a galaxy can be found in Treuthardt et al. (2008). 

The best fitting $\mathcal{R}$ of each galaxy was determined by visually 
comparing the observed $B$-band morphology to that produced in a series of 
models.  Deep $B$-band images with good resolution are ideal for comparing 
observed low velocity dispersion, young stellar components, such as resonance rings, to our models. 
In our simulations, the viscosity within various regions of the 
gas particle system may not be comparable to real systems, meaning features seen in the simulation snapshots can correspond to 
different physical timescales in different regions.  With that in mind, we primarily compared the morphology in the region near the ends 
of the bar (such as the size and shape of any inner rings) but also considered 
the outer spiral structure, especially any outer ring features. Since we are dealing with gas particle simulations of barred spiral galaxies, areas within
the bar region of our models tend to be void of gas particles and are not considered in matching with the observed images.  Four 
independent estimates of the model morphology that best matches the observed
morphology were used to determine the weighted average value and 1$\sigma$ error of $\mathcal{R}$ given in Table 1. In some cases
a slower $\Omega_{p}$ provided a better fit to the outer spiral structure while
a faster $\Omega_{p}$ provided a better fit to the inner structure. This is 
not unexpected since N-body simulations have shown that the spiral component
may have a slower pattern speed then that of the bar (e.g. Rautiainen \& Salo 
1999). This may be the reason why we tend to find few galaxies in our sample 
with convincingly fast bars.  Figure 2 shows the deprojected $B$-band and best 
model images corresponding to our estimated $\mathcal{R}$. Subsection $2.3.1$ 
provides brief discussions on the individual galaxies and the 
corresponding models.  

A number of our sample galaxies appear to have dust lanes along the bar 
major axis when examined in the high resolution $B$-band images from CGS. 
We make a note 
of these dust lanes below in the comments on individual galaxies. 
Athanassoula (1992) has experimented with two-dimensional time-dependent 
hydrodynamical 
simulations that produce dust lanes in barred spiral galaxies. The gravitational
potentials used in the simulations consist of two axisymmetric components, 
which approximate the bulge and disk potential, and a Ferrers ellipsoid,
which approximates the bar potential. A dark halo component was not included 
in the models. Based on the dust lane morphology observed in her models, 
Athanassoula concludes that bars containing dust lanes must be in the fast
domain ($\mathcal{R}=1.2 \pm 0.2$). She claims the convex curvature 
of dust lanes shown in her slow bar models are not observed in real galaxies 
and concludes that slow bars do not exist. The diffuse nature of the few dust 
lanes seen in our sample of galaxies generally makes the morphology of the 
dust lanes difficult to determine visually.
 
\begin{figure*}
\includegraphics{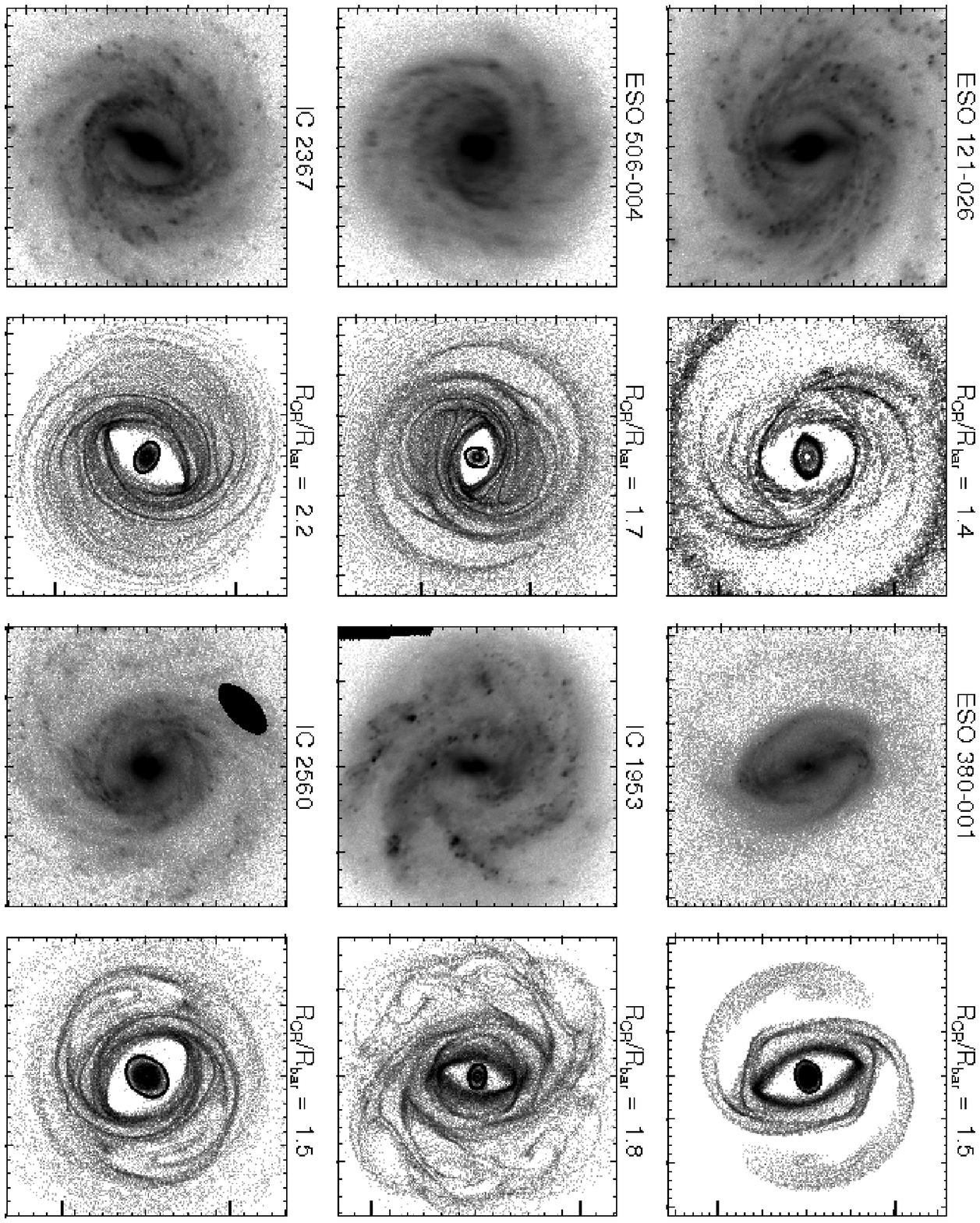}
\vspace*{8.5cm}
\label{landfig}
\end{figure*} 

\begin{figure*}
\includegraphics{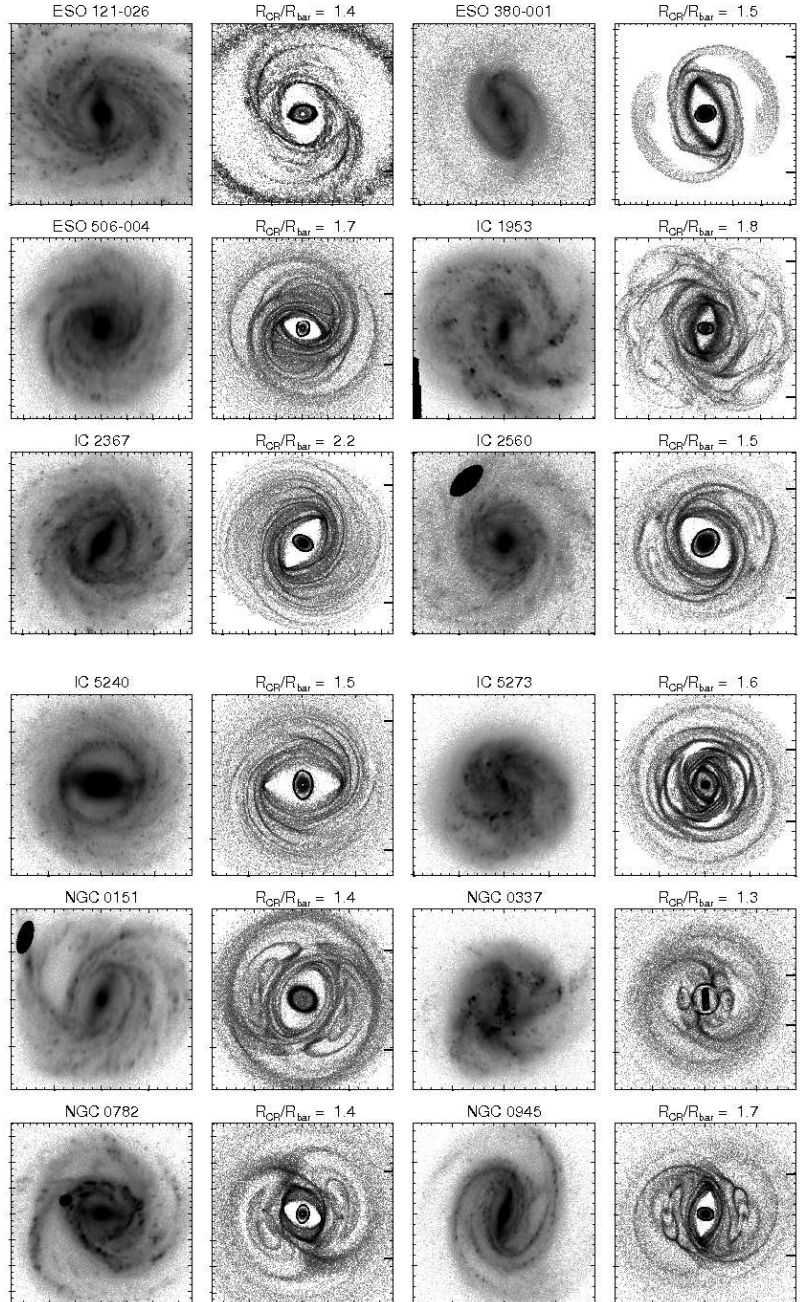}
\vspace*{8.5cm}
\caption{Deprojected $B$-band and best gas particle model images of our sample 
of galaxies. The models shown correspond to our estimated values of
$\mathcal{R}$. Some $B$-band images display black masks that were used when a 
foreground star could not be otherwise removed. The thick tick-marks along the 
right edge of the gas particle images indicate the diameter of corotation.}
\label{landfig}
\end{figure*}

\begin{figure*}
\includegraphics{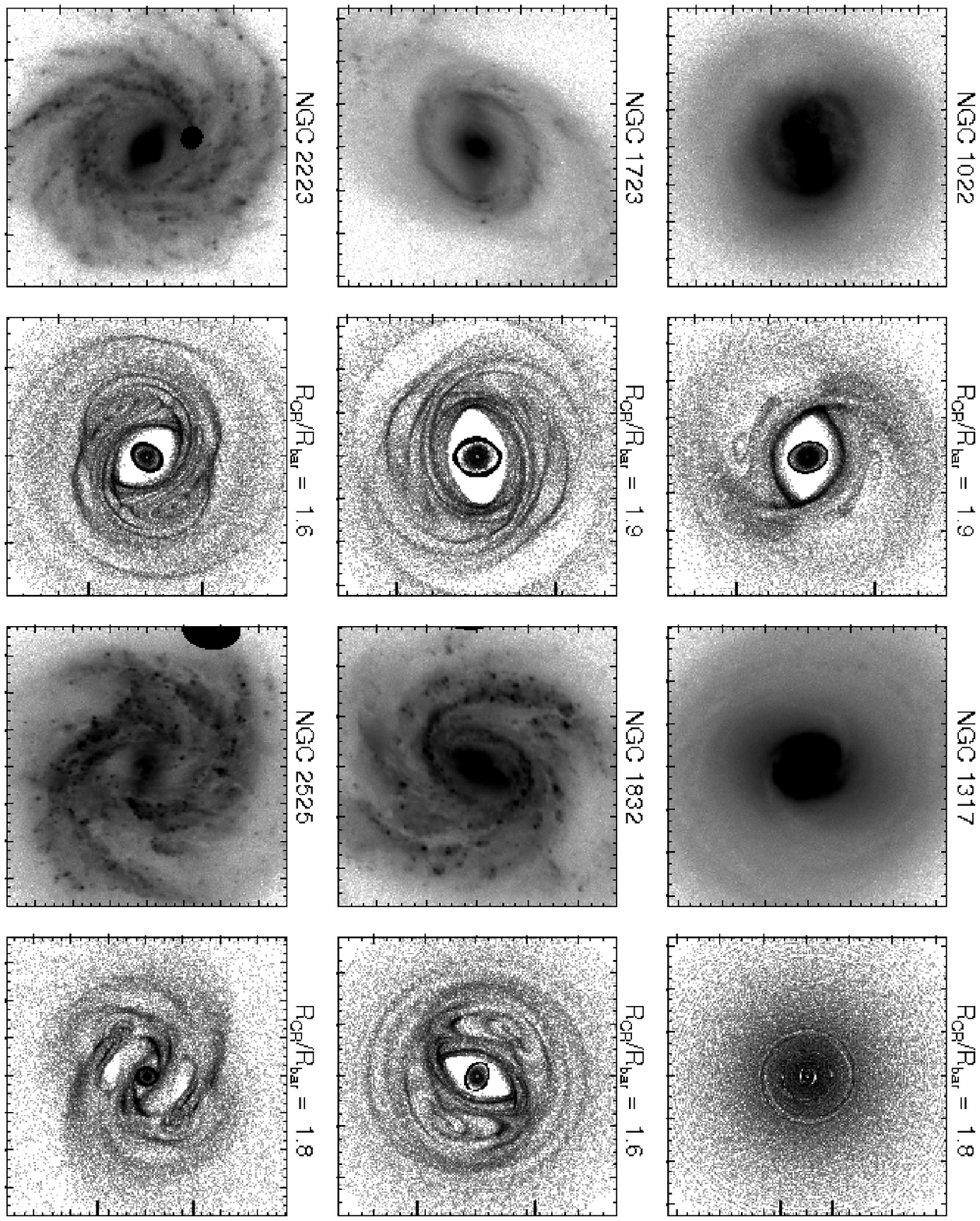}
\vspace*{8.5cm}
\label{landfig}
\end{figure*} 

\begin{figure*}
\includegraphics{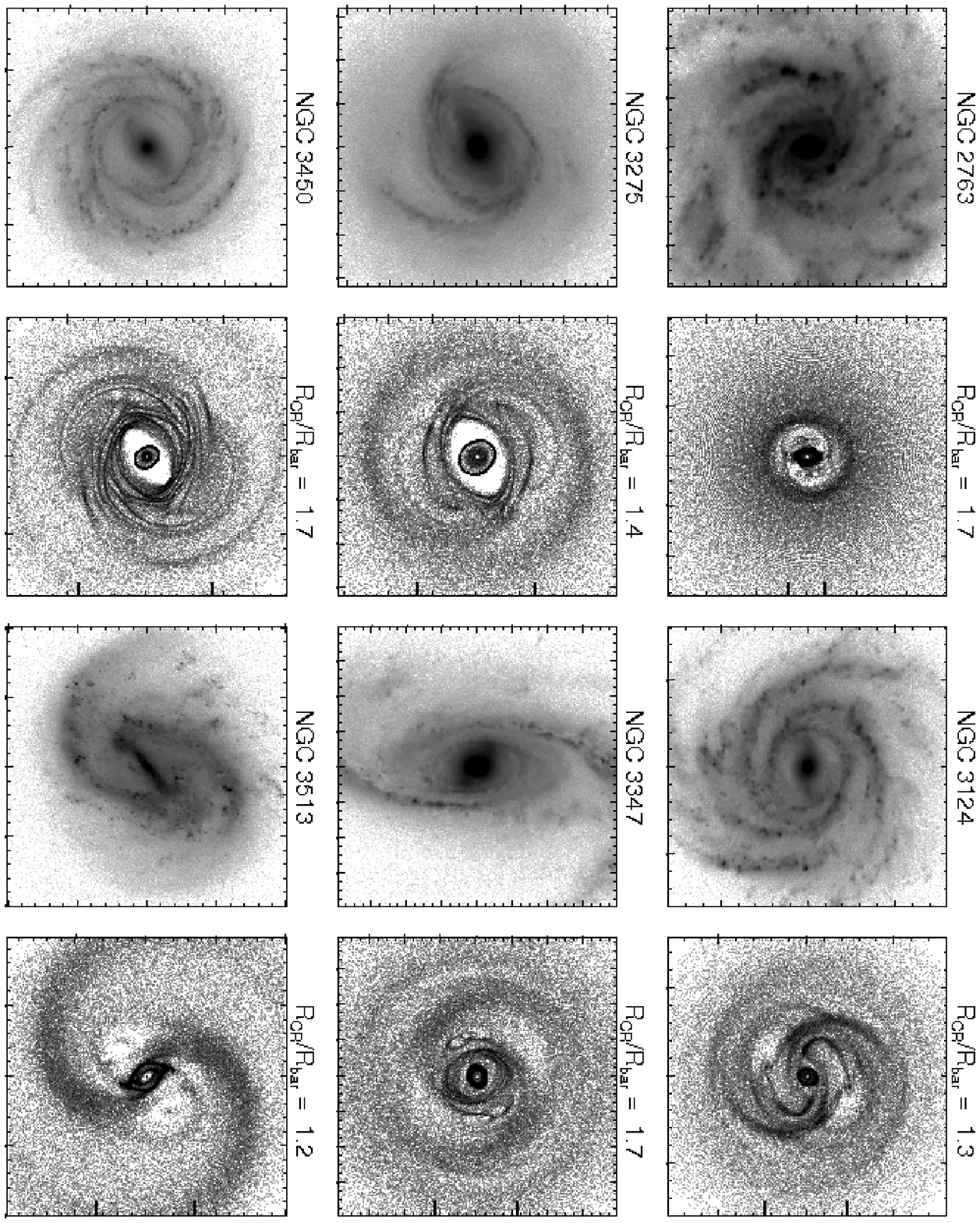}
\vspace*{8.5cm}
\label{landfig}
\end{figure*} 

\begin{figure*}
\includegraphics{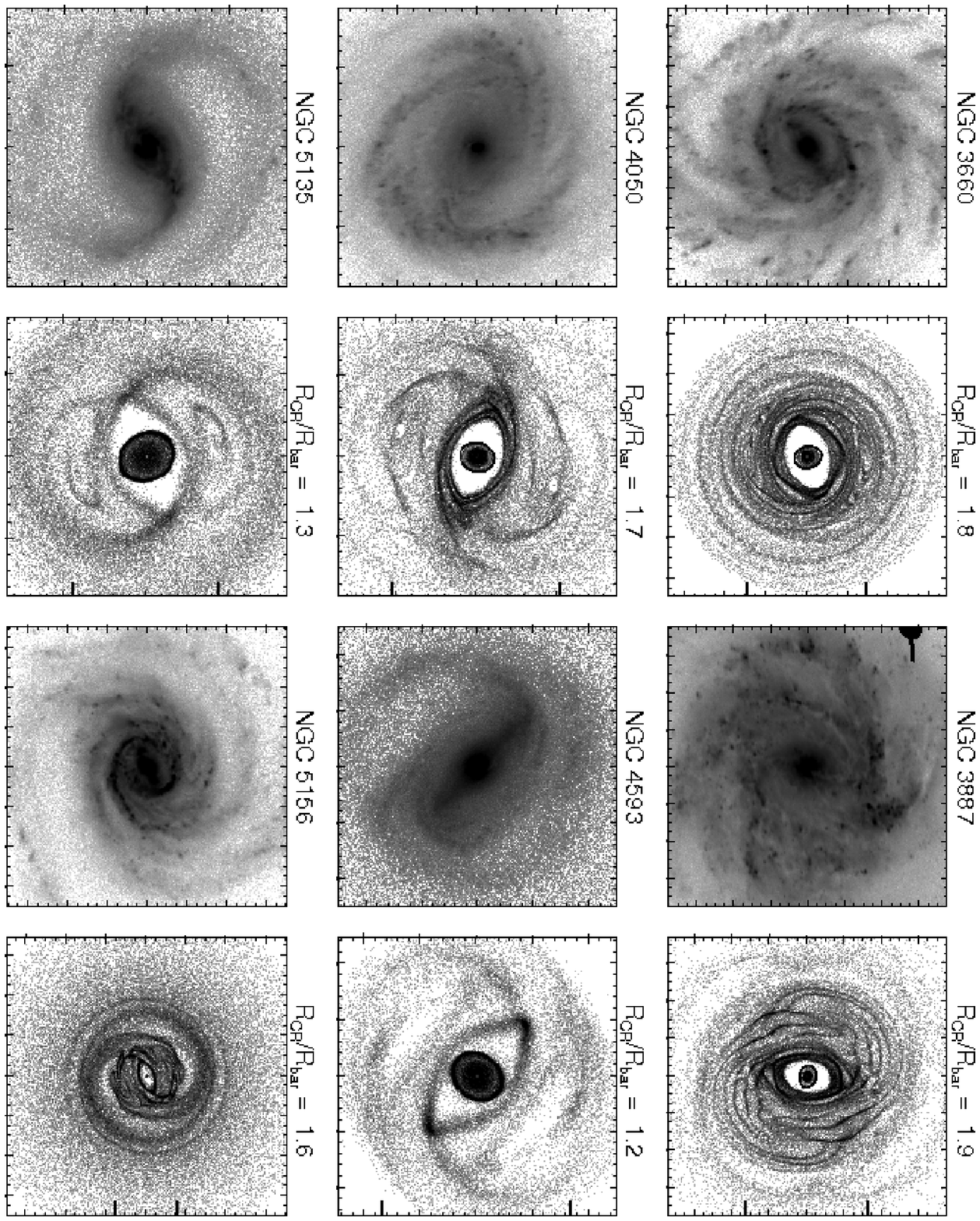}
\vspace*{8.5cm}
\label{landfig}
\end{figure*} 

\begin{figure*}
\includegraphics{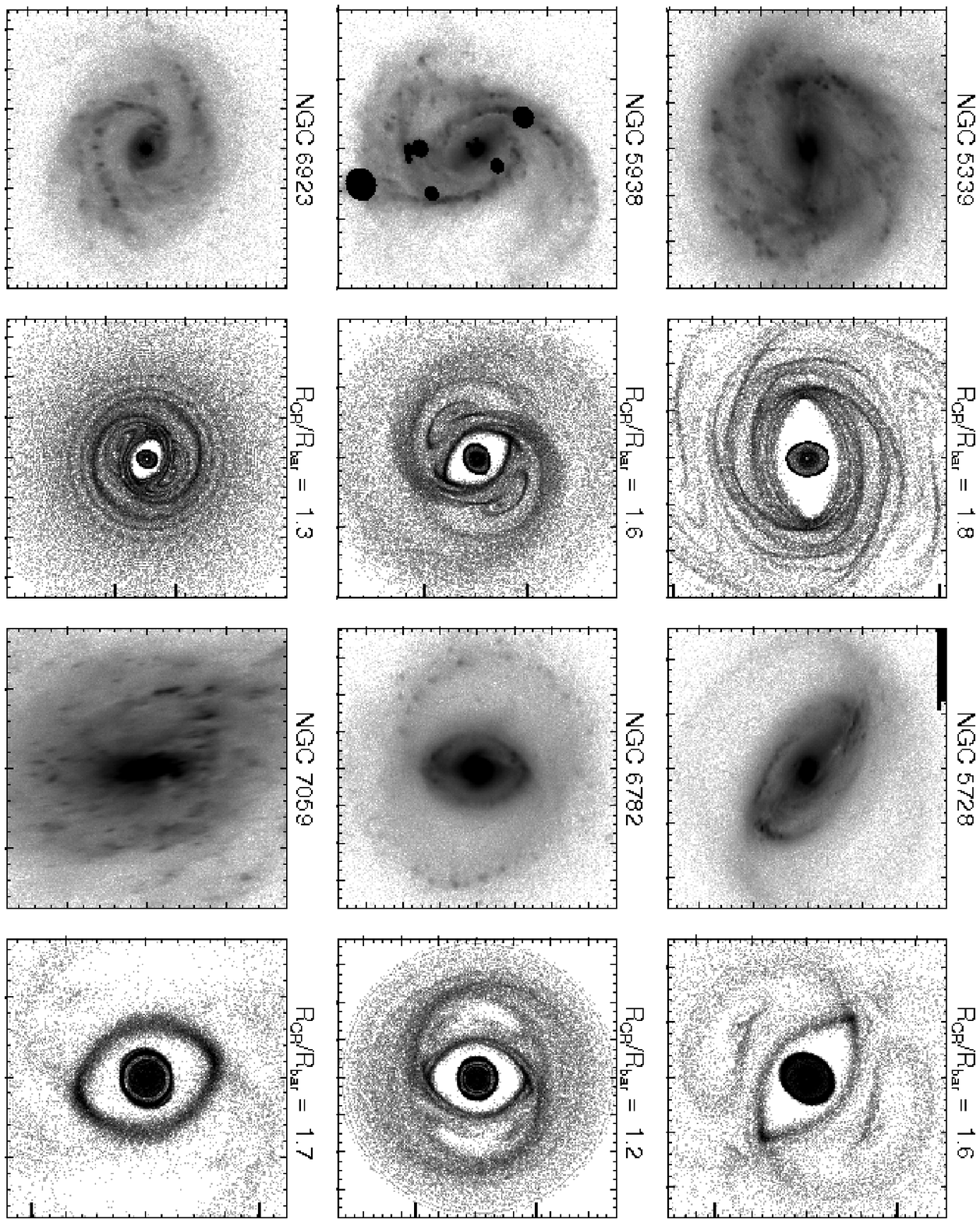}
\vspace*{8.5cm}
\label{landfig}
\end{figure*} 

\begin{figure*}
\includegraphics{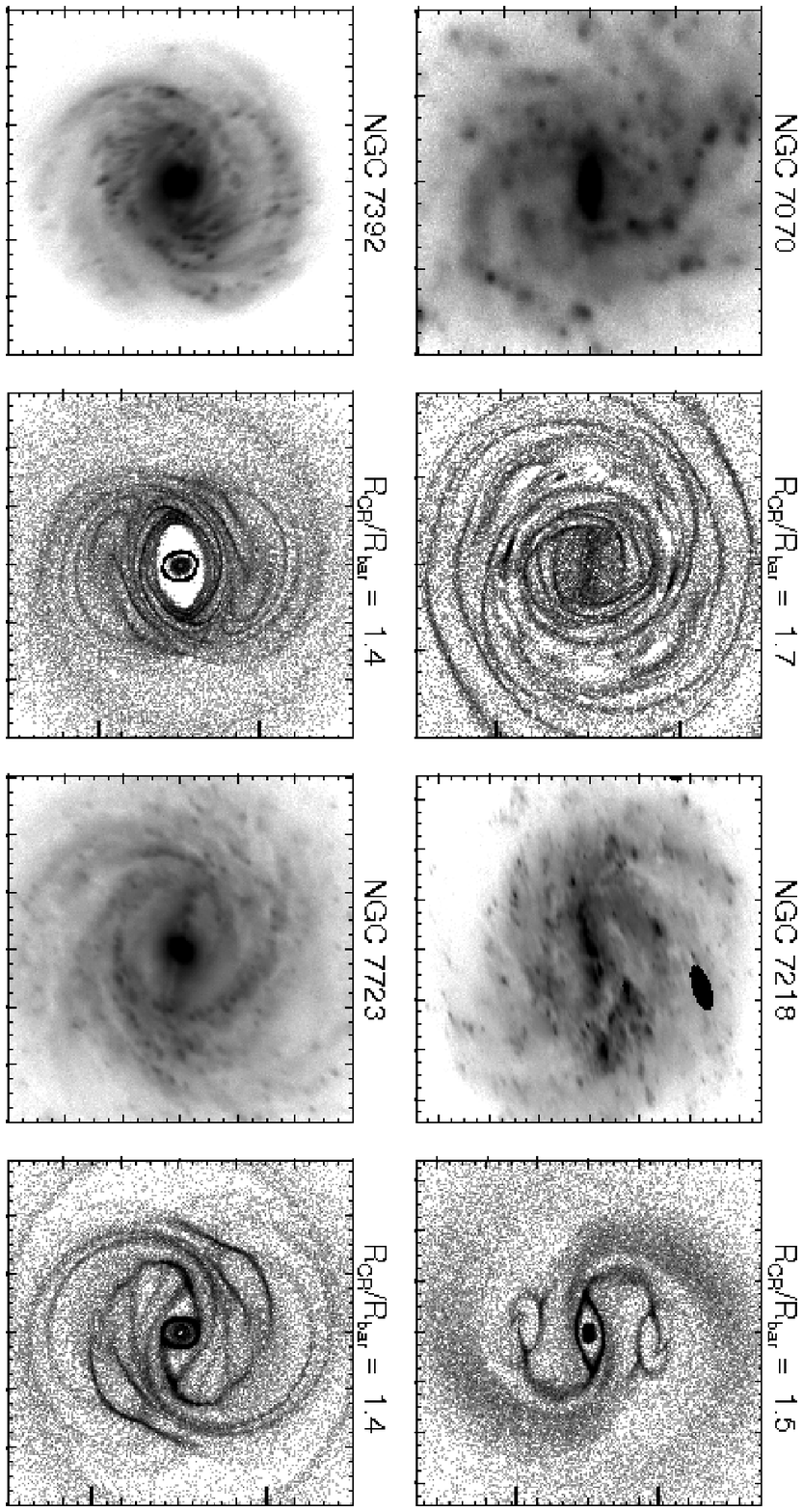}
\vspace*{5.5cm}
\label{landfig}
\end{figure*} 

\subsubsection{Comments on Individual Galaxies}

\emph{ESO 121-026}: Our best models were not able to reproduce the outer 
flocculent arms of this SB(rs)bc galaxy (de Vaucouleurs et al.\ 1991). The 
inner ring, however, is well matched by the model shown in Figure 2. The model
also displays a pointy ring elongated perpendicularly to the major axis and 
encircling the inner ring. This pointy ring is also seen in the observed image.
The $B$-band image displays a possible diffuse dust lane extending along only 
the bottom half of the bar's major axis.

\emph{ESO 380-001}: The inner pseudoring and faint outer ring of this galaxy,
seen in the $B$-band image, is reflected in the (R$^{\prime}$)SB(s)b 
classification by de Vaucouleurs et al.\ (1991). In the observed image, the 
spiral arms can be traced from their origin at the ends of the bar, through
their clockwise winding around the bar, and finally transitioning into the 
outer ring. Our models display a similar morphology but also include an obvious
inner ring.  This may be hinted at in the $B$-band near the bar end 
closest to the bottom of the image. Symmetric dust lanes are also seen extending
along the major axis of the bar in the $B$-band image.

\emph{ESO 506-004}: The model shown for this SAB(r)ab galaxy (de Vaucouleurs 
et al.\ 1991) displays a good match to the observed inner ring and the strong 
spiral arm originating at the left end of the inner ring major axis.

\emph{IC 1953}: Our models of this SB(rs)d galaxy (de Vaucouleurs et al.\ 1991)
produce a series of nested rings surrounded by outer flocculent arms.  The 
outermost modeled ring is a hexagonal 
structure whose major axis appears rotated slightly counter-clockwise to the 
bar major axis.  Hints of this outer most ring structure along the major axis
is seen in the $B$-band image. Within this elongated hexagonal ring is a nearly
circular ring surrounding a ring elongated along the bar major axis. Hints of 
this nearly circular ring, but not the inner elongated ring, are also present 
in the observed image.

\emph{IC 2367}: The size and extent of the inner ring of this SB(r)b galaxy 
(de Vaucouleurs et al.\ 1991) is well matched by the model shown.

\emph{IC 2560}: The inner ring of this (R$^{\prime}$)SB(r)b galaxy (de 
Vaucouleurs et al.\ 1991) is well reproduced by our best estimate model. Also,
there appear to be dust lanes running along the major axis of the bar when 
viewed in the $B$-band.

\emph{IC 5240}: Buta et al.\ (2007) classify this galaxy as SB(r)a with a clear
inner ring and flocculent exterior spiral structure.  Upon closer inspection, 
the upper half of the inner ring appears to consist of three straight arm 
segments. Our simulations fail to produce models that 
directly recreate the strong, nearly circular inner ring seen in the 
deprojected $B$-band image.  Our best model does approximate the inner ring 
morphology from a series of disjointed spiral arms. There appears to be a dust
lane along only the right half of the bar's major axis which extends into dust
in the inner ring.
 
\emph{IC 5273}: The $B$-band image of this SB(rs)cd galaxy (de Vaucouleurs et 
al.\ 1991) does not display any obvious resonance features such as rings. We
attempted to match the inner spiral arms, near the bar semimajor axis length, 
of our models to that in the observed image.  For example, the observed spiral
 arm originating near the bottom end of the bar major axis is matched in our 
given model. 

\emph{NGC 0151}: Our models are not able to recreate the 
asymmetry observed in the outer spiral structure of this SB(rs)bc galaxy 
(Buta et al.\ 2007) since only the even Fourier $m$ values were used to 
calculate the disk gravity. This causes our models to be bisymmetric. Our 
estimate of $\mathcal{R}$ came from matching the the observed and modeled 
inner ring morphology.

\emph{NGC 0337}: This SB(s)d galaxy (de Vaucouleurs et al.\ 1991) has the 
strongest bar of our sample ($Q_{g} = 0.800$). Our models produce a small inner
ring not visible in the $B$-band image.  Our estimates of $\mathcal{R}$ were 
based on comparing the modeled $L4$ and $L5$ Lagrangian regions and the 
morphology of 
the spiral arms extending from the inner ring to that in the observed image.

\emph{NGC 0782}: This SB(r)b galaxy (de Vaucouleurs et al.\ 1991) appears to be
a three armed system when observed in the $B$-band. As was discussed regarding
NGC 0151, our models are necessarily
bisymmetric and will not produce three armed systems. Our 
best estimate of $\mathcal{R}$ came from attempting to match the inner rings 
and the arm structure originating at the bottom of the inner ring in both the 
observed and modeled image.     

\emph{NGC 0945}: de Vaucouleurs et al.\ (1991) classify this galaxy as SB(rs)c.
Our models are able to reproduce the shape and size of the observed inner ring 
but the outer spiral structure is more open than seen in the observed image. 
This may be the result of NGC 0948 which is visually nearby and classified as 
having an uncertain SB(s)c morphology (de Vaucouleurs et al.\ 1991). NGC 0945
also appears to have asymmetrical dust lanes running along the length of the 
bar major axis.

\emph{NGC 1022}: Buta et al.\ (2007) classify this galaxy as a peculiar 
(R)SB(\underline{r}s)a. Our model, corresponding to our best estimate of 
$\mathcal{R}$, provides a good match to both the inner and outer ring 
morphology. 

\emph{NGC 1317}: Buta et al.\ (2007) describe this galaxy as having a
double-barred morphology and classify it as (R$^{\prime}$)SAB(rl)a. We attempted
to recreate the observed morphology by varying $\Omega_{p}$ of the nuclear bar,
which is the weakest in our sample ($Q_{g} = 0.040$),
rather than the less prominent broad oval, or primary bar. Our models produced 
a tight spiral structure around the nuclear bar with a higher
density of gas particles on the leading edge of the spiral and a lower density
on the trailing edge. We matched the lower density trailing edge 
of our models to the gaps seen in the $B$-band emission surrounding the 
nuclear bar region. Sticky-particle modeling of this galaxy by 
Rautiainen et al. (2008) led to an estimate of 
$\mathcal{R} = 0.89 \pm 0.25$, nearly half of our estimate. Although Rautiainen
et al.\ (2008) do not comment on this specific galaxy, it is apparent from 
Figure 4 of their paper that their models were produced by varying $\Omega_{p}$
of the broad oval rather than the nuclear bar. The nuclear bar does not display
any apparent dust lanes, but a nearly symmetrical dust spiral is seen in the 
broad oval region. 

\emph{NGC 1723}: This peculiar SB(r)a galaxy (de Vaucouleurs et al.\ 1991) may 
be the product of an interaction with a nearby galaxy. The fact that one of the
outer spiral arms extends and curves toward a smaller field galaxy in our raw 
$B$-band image is apparent evidence for this.  Our simulations were not able to 
recreate the observed outer spiral structure or the slightly oblong nature of 
the inner ring seen in our deprojected image. The model corresponding to our 
weighted mean value of $\mathcal{R}$ contains an inner ring of similar size to 
the observed inner ring. Larger or smaller values of $\mathcal{R}$ result in 
models with larger or smaller inner rings, respectively.

\emph{NGC 1832}: de Vaucouleurs et al.\ (1991) classify this galaxy as SB(r)bc.
Our model, corresponding to our weighted mean estimate of $\mathcal{R}$, 
produces an inner ring of similar extent to that seen in the observed image.
The inner ring in the $B$-band image may be the result of gas circulating along
so called ``banana'' orbits around the $L4$ and $L5$ Lagrangian points. This 
is also seen in our model, but the density of gas particles is not very large 
in the portions of these orbits that bring them close to the bar. Rautiainen 
et al.\ (2008) estimate $\mathcal{R}$ to be $1.74 \pm 0.40$ from their 
sticky-particle simulations using the same code as the current simulations. 
Within the errors, this coincides with our estimate of 
$\mathcal{R} = 1.61 \pm 0.08$.

\emph{NGC 2223}: Our model of this SAB(r)b galaxy (de Vaucouleurs et al.\ 1991)
produces a strong inner ring of the same size and extent as the one that is 
hinted at in $B$-band image. Our model also recreates the inner spiral arm 
structure observed just beyond the ends of the inner ring major axis.

\emph{NGC 2525}: This SB(s)c galaxy (de Vaucouleurs et al.\ 1991) is well 
matched by our model. The size and position of the modeled $L4$ and $L5$ 
Lagrange regions duplicate that seen in the $B$-band image. The modeled spiral 
structure also closely matches the observations, although the gas particle 
density is not high.

\emph{NGC 2763}: de Vaucouleurs et al.\ (1991) classify this galaxy as a 
peculiar SB(r)cd. The simulation model of this galaxy, shown in Figure 2, is
the weighted mean of individually estimated $\mathcal{R}$ values ranging from 
$1.2 \pm 0.2$ to $1.9 \pm 0.2$.  

\emph{NGC 3124}: Buta et al.\ (2007) classify this galaxy as SB(r)bc. This 
galaxy does
not host a traditional bar but rather a bar-spiral hybrid winding in the 
opposite sense as the outer spiral arms. This is especially apparent in the
$K_{s}$-band image.  Our models do not reproduce the nearly circular looking 
inner ring seen in the $B$-band but rather approximate it as a pair of 
disjointed spiral arms. A one-armed dust spiral is seen in the $B$-band
winding in the same sense as the stellar spiral arms. This one-armed dust 
spiral is seen crossing the left semi-major axis of the bar. 

\emph{NGC 3275}: The apparent inner ring seen in the $B$-band image of this 
SB(r)ab galaxy (Buta et al.\ 2007) is recreated by our best estimate model in
the form of spiral arm segments outside of a true inner ring. A faint hint of
a thin ``true'' inner ring is seen in the observed image near the left end of 
the bar, inside the apparent inner ring. A morphological match is also seen 
when comparing the outer spiral arm that originates from near the right end of 
the bar. Models produced by Rautiainen et al.\ (2008) estimate 
$\mathcal{R} = 1.53 \pm 0.34$ which coincides with our 
estimated value within the errors. The $B$-band image of this galaxy shows 
diffuse, nearly symmetrical dust lanes and spirals in the bar region. The 
dust lanes are originate near the ends of the bar and transition into spirals 
at the bar ends.

\emph{NGC 3347}: Our assumption of intrinsically circular galaxy disks
does not appear to be valid for this galaxy which leads to a somewhat 
unrealistic deprojected image. NGC 3347 is a member of a group 
(Garcia 1993) and appears, in the CGS images, to be interacting with a 
nearby peculiar spiral galaxy of uncertain morphology (NGC 3354; de 
Vaucouleurs et al.\ 1991). Our simulation models were not able to recreate the 
tightly 
wrapped outer spiral arms of this SB(rs)b galaxy (de Vaucouleurs et al.\ 1991).
The shape and extent of the modeled inner ring closely matches that surrounding
the observed bar.

\emph{NGC 3450}: The observed morphology of this SB(\underline{r}s)b galaxy 
(Buta et al.\ 2007) is well matched by our model. The inner ring, as well as the
outer spiral structure, appear to agree in both the observed and modeled image. 
A dust lane elongated nearly perpendicular to the semi-major axis of the lower 
left end of the bar is seen in the $B$-band image.

\emph{NGC 3513}: This galaxy is classified as SB(s)c by Buta et al.\ (2007) and 
contains one of the strongest bars of our sample ($Q_{g} = 0.540$; see Table 1).
Our 
simulations were unable to easily reproduce the main spiral arms alone. 
Instead, most of our gas particle models show diffuse arms emanating 
perpendicularly to the 
bar from an apparent nuclear ring. In the model associated with our weighted
average estimate of $\mathcal{R}$, a smaller number of gas particles also 
appear
to emanate from this ring, run parallel to the bar, curve around the bar ends, 
and finally reconnect to the stronger, more diffuse arms on the opposite side 
of the ring. This less prominant gas particle distribution more closely 
resembles the 
observed morphology than the more obvious gas particle distribution. 
Sticky-particle modeling by Rautiainen et al.\ (2008) produces an estimate of 
$\mathcal{R} = 1.50 \pm 0.29$. Our estimate of $\mathcal{R} = 1.24 \pm 0.07$ 
falls within the error of Rautiainen et al.'s estimate.

\emph{NGC 3660}: Recreating the outer spiral arm morphology of this SB(r)bc 
galaxy (Buta et al.\ 2007) proved to be difficult. Our weighted mean estimate
of $\Omega_{p}$, or $\mathcal{R}$, provides a good morphological match of the 
inner ring between both the observation and the model. The $B$-band image shows
faint, symmetrical, and diffuse dust lanes running along the major axis of the 
bar.

\emph{NGC 3887}: Our models of this SB(rs)bc galaxy (Buta et al.\ 2007) produce 
an inner ring that is not visible in the $B$-band image. The extent the inner 
ring shown in our model does correspond to the extent of the straight portion 
of the symmetrical dust lanes seen along the major axis of the bar.  The 
turn-over radius of these dust lanes is similar to the inner ring semimajor 
axis length. The outer spiral structure also approximates the observed spiral 
structure near the bar.

\emph{NGC 4050}: The morphology of this SB(r)ab galaxy (de Vaucouleurs et al.\ 
1991) is well matched by our model.  Both the inner ring and the outer spiral 
structure appear to agree. A single diffuse dust lane is seen in the $B$-band
running from the nucleus along the right bar semi-major axis. 

\emph{NGC 4593}: Buta et al.\ (2007) classify this galaxy as 
(R$^{\prime}$)SB(rs)ab. Our model does well to approximate the shape and extent 
of the inner and outer ring. The outer ring appears to be due to gas trapped 
in ``banana'' orbits and circulating around the $L4$ and $L5$ Lagrangian 
regions, as was seen in the simulation models of NGC 1433 (Treuthardt et al. 
2008). The observed image of NGC 4593 also shows an apparent ``plume'', 
similar to those seen in NGC 1433, near the leading edge of the bar and in the 
upper left quadrant of the $B$-band image.  Our model also produces a linear 
component to the ``banana'' orbits, close to the outer boundary of the inner 
ring. This feature is also seen in lower half of the observed image. A strong 
dust lane is seen along the upper left semi-major axis of the bar, while a 
fainter, more diffuse dust lane is seen along the opposite semi-major axis.

\emph{NGC 5135}: This SB(s)ab galaxy (de Vaucouleurs et al.\ 1991) was well
matched by our model.  Both the inner ring and outer spiral morphology are 
reproduced. The inner regions of this galaxy display a lot of dust in the 
$B$-band, but no dust lanes are apparent.

\emph{NGC 5156}: We were not able to reproduce the overall outer spiral 
structure observed in this SB(r)b galaxy (de Vaucouleurs et al.\ 1991). 
Determining a good morphological match to the inner ring also proved to be 
difficult as our estimates of $\mathcal{R}$ ranged from $1.4 \pm 0.2$ to 
$1.8 \pm 0.2$. The model we show, corresponding to the weighted mean of our 
$\mathcal{R}$ values, does approximate the two observed spiral arms closest to
the bar, in the lower half of the image.

\emph{NGC 5339}: This peculiar SB(rs)a galaxy (de Vaucouleurs et al.\ 1991) is 
well matched by our model.  The outer spiral arm morphology is recreated by a
series of arms that form a slight hexagonal pattern. The extent and shape of the
inner ring is also approximated. Nearly symmetrical dust lanes are seen in the 
$B$-band along the major axis of the bar.

\emph{NGC 5728}: The inner ring of this (R$_{1}$)SA\underline{B}(r)a galaxy 
(Buta et al.\ 2007) is reproduced by our model. An outer ring is also created 
by our simulation, but the observed ring is much more circular in the 
deprojected image. Symmetrical dust lanes are also seen along the bar major axis
when viewed in the $B$-band.

\emph{NGC 5938}: A large number of foreground stars are in the field of this 
SB(rs)bc galaxy (de Vaucouleurs et al.\ 1991). Those stars that could not be 
blended to the background where masked, leaving obvious artifacts in the image.
The spiral structure far from the bar is not well matched by our model.  The
left side of the observed apparent inner ring is matched by our modeled inner 
ring. The right side of the observed apparent inner ring is better matched by 
the modeled spiral arm originating just outside the inner ring.  

\emph{NGC 6782}: This galaxy is classified as (R$_{1}$R$_{2}^{\prime}$)SB(r)a 
by Buta 
et al.\ (2007). The outer ring features are best fit by $\mathcal{R} 
\approx 1.3$ while the inner ring is best fit by $\mathcal{R} \approx 1.1$. In
either case, the bar is still classified as fast. For this galaxy, 
Rautiainen et al.\ (2008) 
estimate $\mathcal{R} = 1.25 \pm 0.23$ but they  were not able to recreate 
the outer pseudo-ring in their models. Our models display this outer pseudo-ring
with a very similar value of $\mathcal{R}$. Lin et al.\ (2008) produced 
two-dimensional hydrodynamical simulations of this galaxy and were able to 
recreate the symmetric dust lanes observed in the $B$-band. They, however, found
$\mathcal{R}$ to be 1.5.

\emph{NGC 6923}: Our model morphology matches the observed inner ring and lower
spiral arm of this SB(rs)b galaxy (de Vaucouleurs et al.\ 1991).

\emph{NGC 7059}: This highly inclined ($i = 71.1^{\circ}$; see Table 1) 
SAB(rs)c (de Vaucouleurs et al.\ 1991) galaxy displays little structure in its 
deprojected $B$-band morphology that 
could be thought of as an unambiguous, organized product of $\Omega_{p}$.  All 
of our 
simulations produce models with a clear inner ring which is not evident in the 
$B$-band.  Many of our models also display four low particle density spiral 
arms extending from near the ends of the major axis of the inner ring. This
spiral structure may, debatably, be hinted at in the deprojected $B$-band 
image.  Consequently, the weighted 
mean of our best morphological matching estimates ($\mathcal{R} \approx 1.7$) 
was 
determined from individual estimates ranging from $\mathcal{R} = 1.9 \pm 0.1$ 
to $\mathcal{R} = 1.2 \pm 0.2$.

\emph{NGC 7070}: de Vaucouleurs et al.\ (1991) classifies this galaxy as a 
non-barred SA(s)cd. Both the $B$ and $K_{s}$-band images show a clear, highly
elliptical structure at the galaxy's center. We attempted to match the observed 
spiral arm morphology, but this galaxy's flocculent 
structure and lack of any obvious resonance features induced a large amount of
error in our $\mathcal{R}$ estimates. The errors of our individual estimates 
ranged from $\approx 12\%-33\%$.    

\emph{NGC 7218}: The morphology of the pair of spiral arms originating from 
the left end of the
observed bar of this SB(rs)cd galaxy (de Vaucouleurs et al.\ 1991) is 
approximated by our model. The more tightly wound of these arms appears to 
connect to a Lagrange region, and continue on to form a wider, more diffuse, and
more open spiral arm in our model. The origin point of this wider arm may be
seen near the upper right end of the bar in the $B$-band image. 

\emph{NGC 7392}: A central oval component of this SA(s)bc galaxy (de 
Vaucouleurs et al.\ 1991) is more apparent in the $K_{s}$-band image than in 
the $B$-band. Our model does well to match the inner pseudo ring, as well as 
the outer spiral structure.  Particularly, our model provides a morphological 
match to a looping ring seen in the upper half of the galaxy image. Faint, 
asymmetrical dust lanes are seen in the $B$-band along the major axis of the 
central oval.
 
\emph{NGC 7723}: Buta et al.\ (2007) classify this galaxy as SB(rs)b. Our model
is able to recreate the inner ring and outer spiral structure from a series of
complete and segmented spiral arms. Modeling by Rautiainen et al.\ (2008) 
produce an estimate of $\mathcal{R} = 1.31 \pm 0.29$ which coincides with our
model-based estimate within the errors. Dust lanes are seen in the $B$-band 
image of this galaxy. The dust lane that runs along the leading edge of the 
right semi-major bar axis is quite apparent and linear. The dust lane running 
along the opposite end of the bar appears to originate from the trailing side 
of the bar and cross to the leading side.

\section{Results and Discussion}

Booth \& Schaye (2010) have developed models of black hole growth in galaxies that appear to reproduce 
the empirical relationship found between black hole mass 
and halo mass (Bandara et al. 2009; see figure 3 of Booth \& Schaye 2010). The implication from 
their models is that galaxies with low dark halo concentrations will host SMBHs with a range of masses, but they will be less massive
than average for a given halo mass (see Figure 3). 

\begin{figure*}
\includegraphics{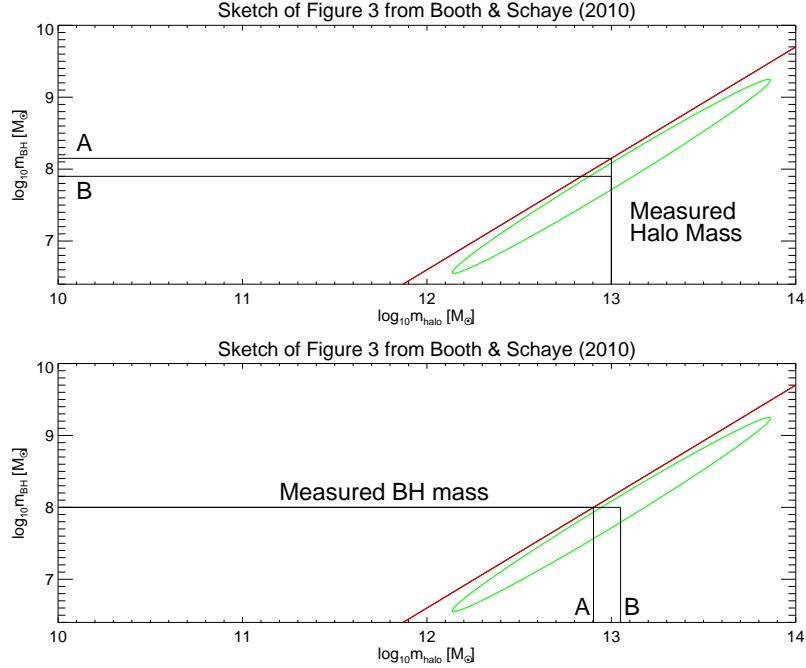}
\vspace*{8.5cm}
\caption{Sketches of Figure 3 from Booth \& Schaye (2010). The red line represents the empirical SMBH and halo mass relationship (Bandara et al. 2009) indicating the average SMBH mass for a given halo mass.
The green ellipse outlines the region of less-than-average SMBH masses due to less concentrated haloes. The upper figure shows that for a galaxy with a low concentration halo and a given measured halo mass, the SMBH mass 
is overestimated (A) when only the empirical relationship is used. In reality, a galaxy with a low concentration halo will have a lower SMBH mass (B) for the same
measured halo mass. The lower figure shows that when the SMBH mass is measured for a galaxy with a low concentration halo, the empirical relationship implies a halo mass that is lower (A) than the actual halo mass (B).}
\label{landfig}
\end{figure*} 

A simple, easily applicable tool for determining the SMBH mass of nearby spiral galaxies is through the relationship 
between $M_{BH}$ and $P$ found by Seigar et al. (2008).
The $M_{BH}-P$ relationship is fit by Seigar et al. (2008) using a double-power-law model given as
\begin{equation}
M_{BH}= 2^{(\beta-\gamma)/\alpha}M_{BH_{b}}\left (\frac{P_{b}}{P}\right)^{\gamma}\left [ 1+\left ( \frac{P}{P_{b}} \right )^{\alpha} \right ]^{(\gamma-\beta)/\alpha},
\end{equation}
where $\beta$ is the slope of the power law for large pitch 
angles, $\gamma$ is the slope of the power law for small pitch angles, $P_{b}$ 
is the transition from small to large pitch angles, $\alpha$ 
governs the sharpness of the transition, and $M_{BH_{b}}$ is the black hole 
mass for a pitch angle $P_{b}$. The values of the parameters that provide this
best fit model are $\beta = 126.1$, $\gamma = 2.92$, $P_{b} = 40.8^{\circ}$, 
$\alpha = 23.5$, and $M_{BH_{b}} = 1.72 \times 10^{4} M_{\sun}$ (see equation 2 of Seigar et al. 2008). Other model fits
have also been suggested such as a linear (Seigar et al. 2008) or cotangent 
relation (Ringermacher \& Mead 2009). 

In Figure 4 (left) we plot our estimated values of $\mathcal{R}$ versus our measurements of pitch angle.  
There does not appear to be any correlation between either fast,
intermediate, or slow bar pattern speed and pitch angle. Those  
galaxies with clearly fast bars span a range in $P$ from 13.4$^{\circ}$ for NGC 6782 to 
32.9$^{\circ}$ for NGC 0337.  The two galaxies with the
smallest pitch angles of those with fast bars are NGC 5135 and NGC 6782. 
The outer spiral arms of these galaxies are not logarithmic as we assume and appear to be the result of gas trapped
in resonance regions associated with $\Omega_{p}$ (see Figure 5). The relative 
amplitudes of the $m=2$ and $m=4$ Fourier components indicate that a single 
spiral pattern is dominate (see Figure 6) meaning that $P$ is not constant with radius in our measurements 
of these two galaxies. When pitch angle is translated to SMBH mass via equation 2 in Seigar et al. (2008) we can see that $M_{BH}$ also does not 
correlate with bar pattern speed (Figure 4, right). That is, galaxies with fast bars, and low central density
dark matter halos (Debattista \& Sellwood 2000), have SMBH masses similar to other barred spiral galaxies.

\begin{figure*}
\includegraphics{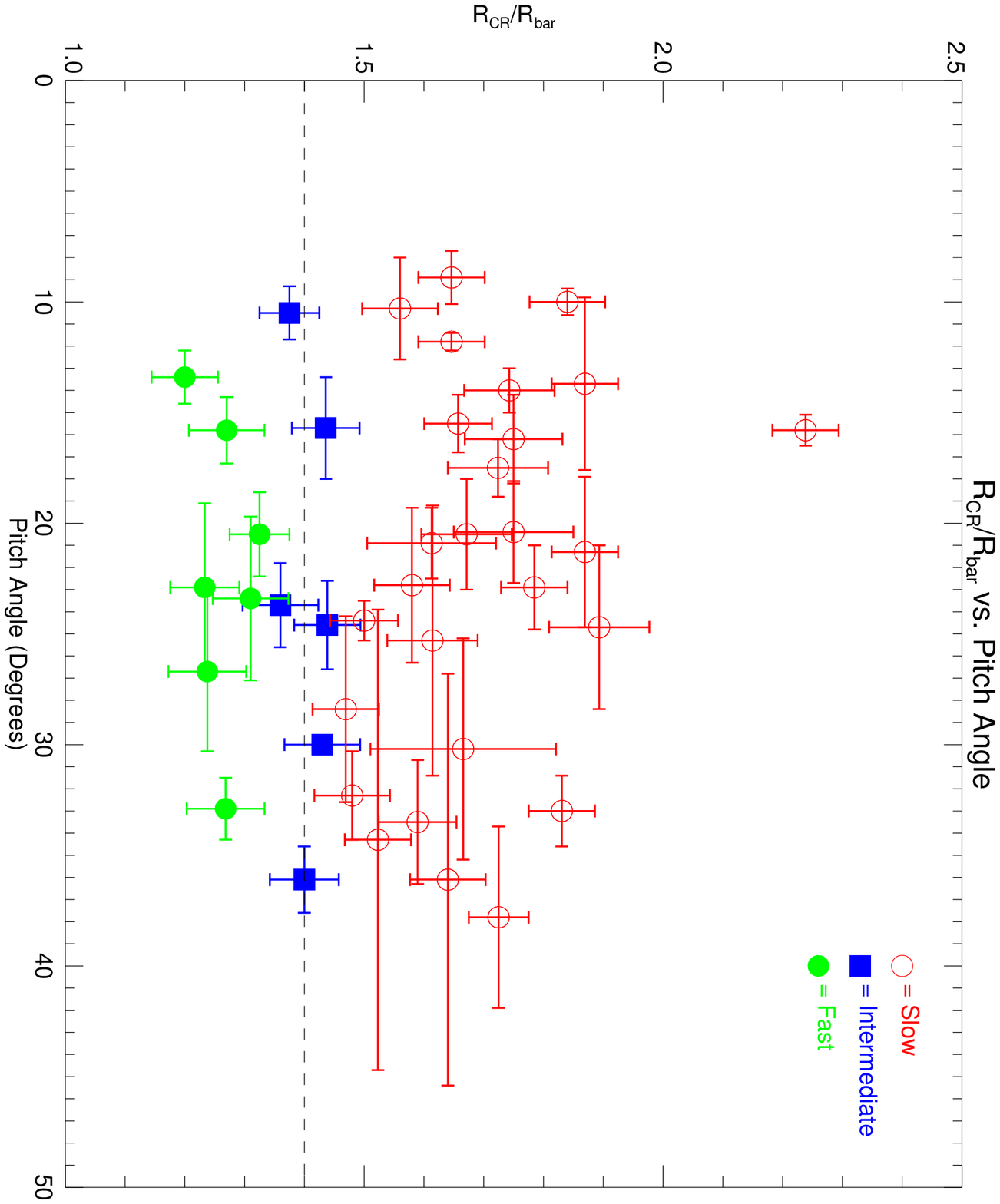}
\includegraphics{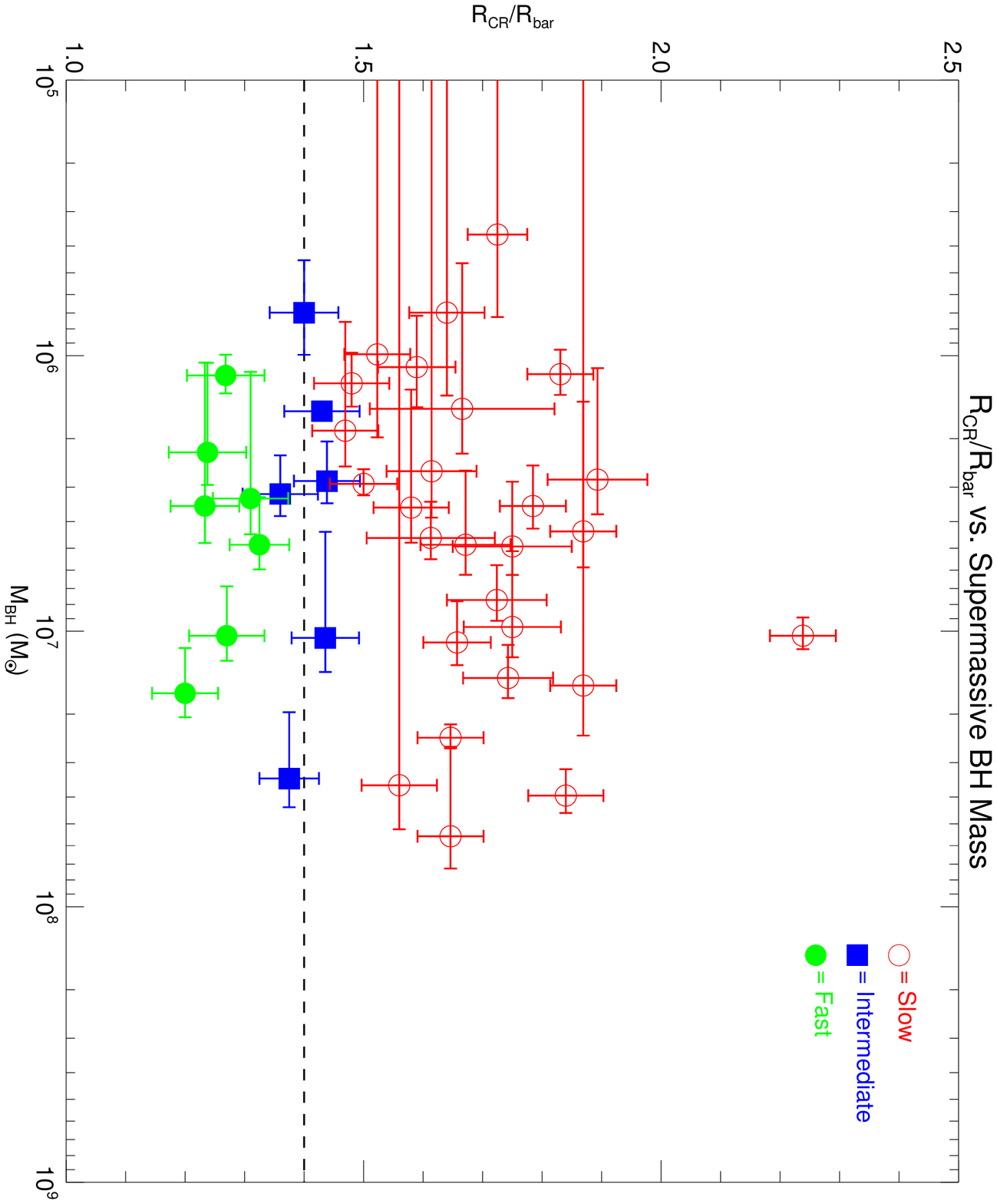}
\vspace*{8.5cm}
\caption{Plots of estimated $\mathcal{R} = R_{CR}/R_{bar}$ versus measured spiral arm pitch angle 
(left) and SMBH mass (right) for galaxies with fast, intermediate,
or slow bar pattern speeds. The red open circles 
correspond to galaxies where $\mathcal{R}$ is estimated to be in the slow bar 
domain within the errors, the green filled circles correspond to galaxies in 
the fast bar domain within the errors, and the blue filled squares correspond 
to galaxies that could fall in either domain given the errors. The fit given by 
equation 2 in Seigar et al. 
(2008) was applied to the pitch angle measurements in the left plot to determine $M_{BH}$ in the right plot. 
The dashed line indicates $\mathcal{R} = 1.4$, the boundry between fast and slow bars. There appears to be 
no correlation between bar pattern speed and pitch angle or black hole mass.}
\label{landfig}
\end{figure*}  

\begin{figure*}
\includegraphics{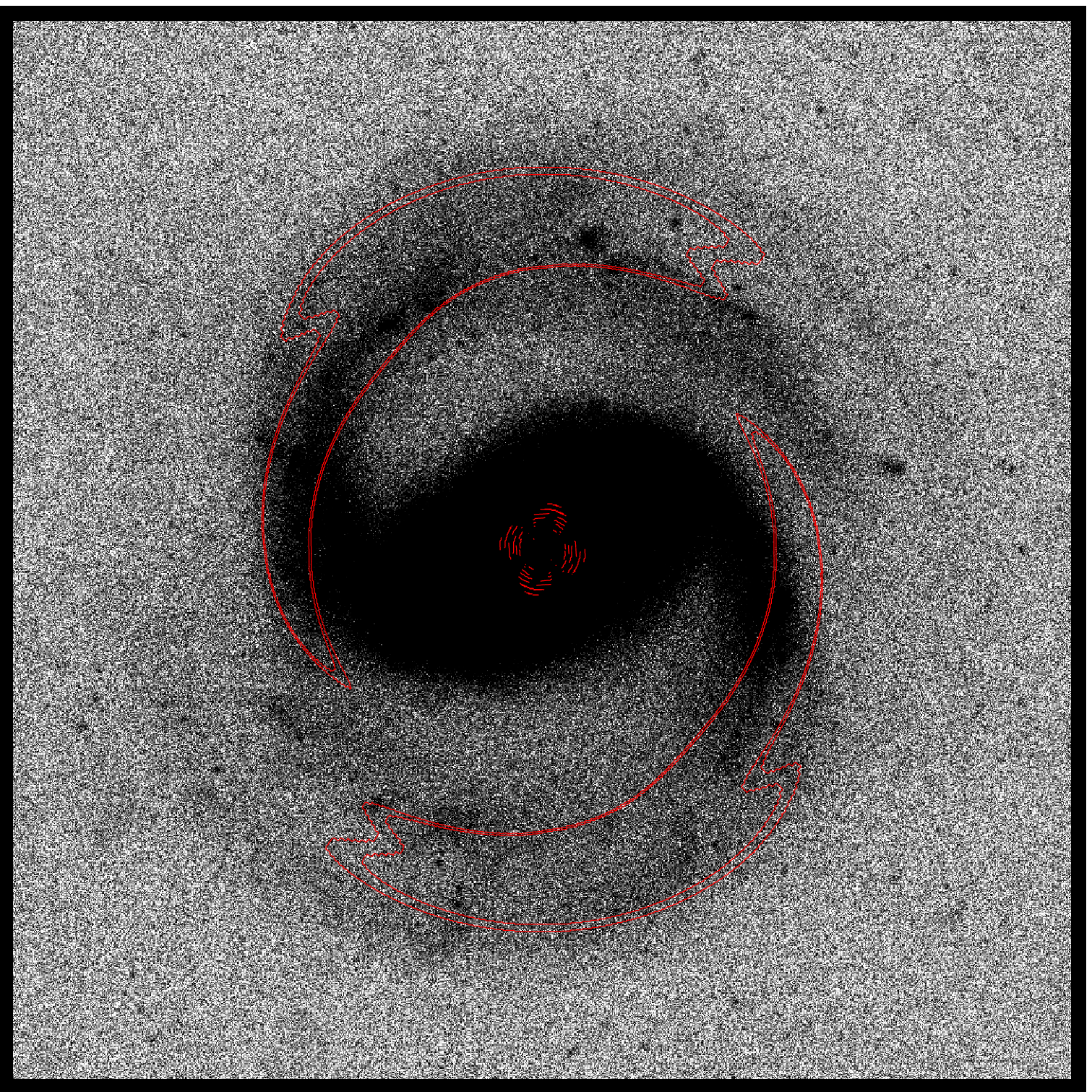}
\includegraphics{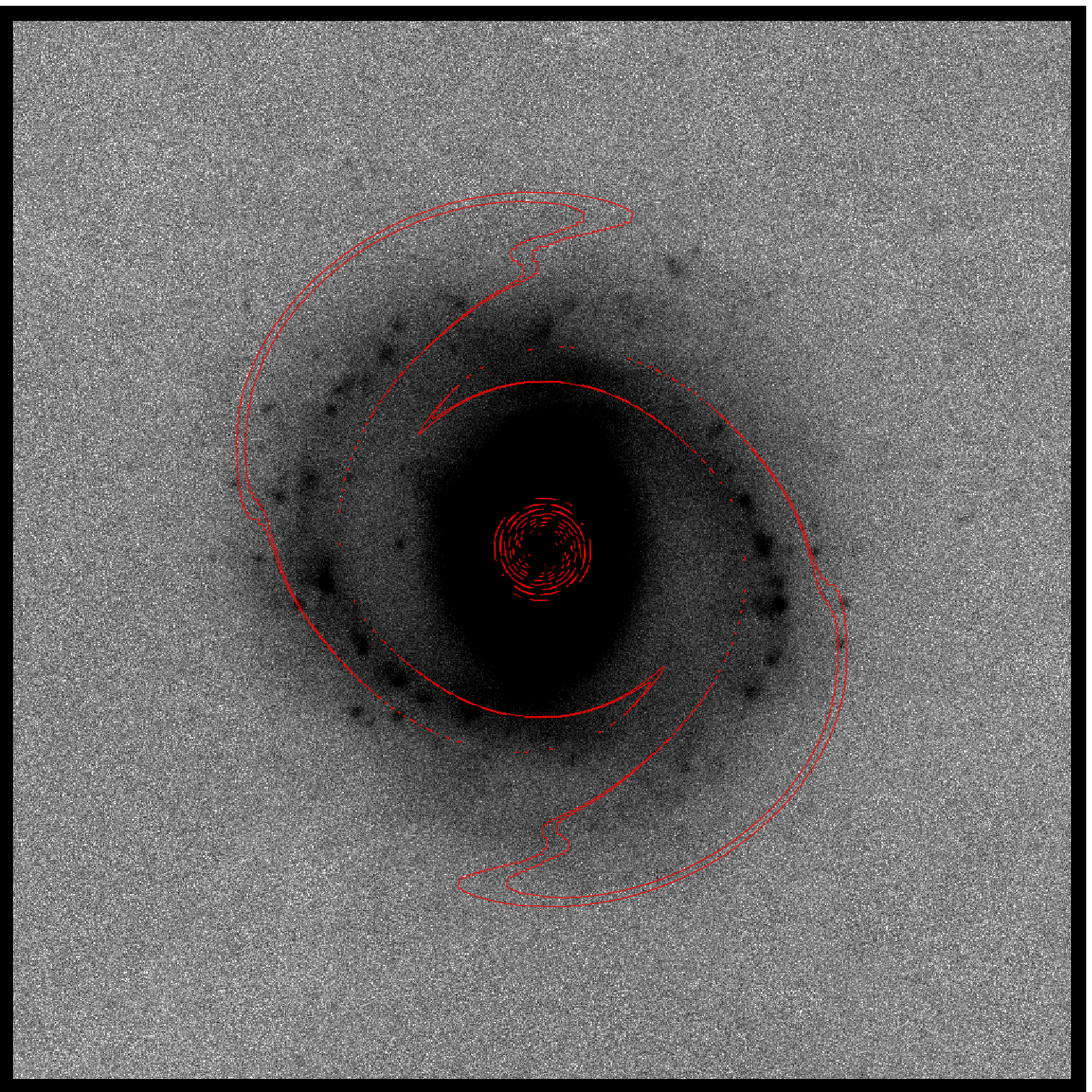}
\vspace*{9.5cm}
\caption{$B$-band images of NGC 5135 (left) and NGC 6782 (right) with contours
of the $m$=2 Fourier components overlayed. The contours are fit to the outer
spiral arms with the assumption that they are logarithmic spirals. 
The figures show that the outer arms appear logarithmic 
at small radii but deviate from this assumption at larger radii. This is 
likely due to gas being trapped at the OLR and implies that $P$ is not 
constant with radius in our measurements.}
\label{landfig}
\end{figure*} 

\begin{figure*}
\includegraphics{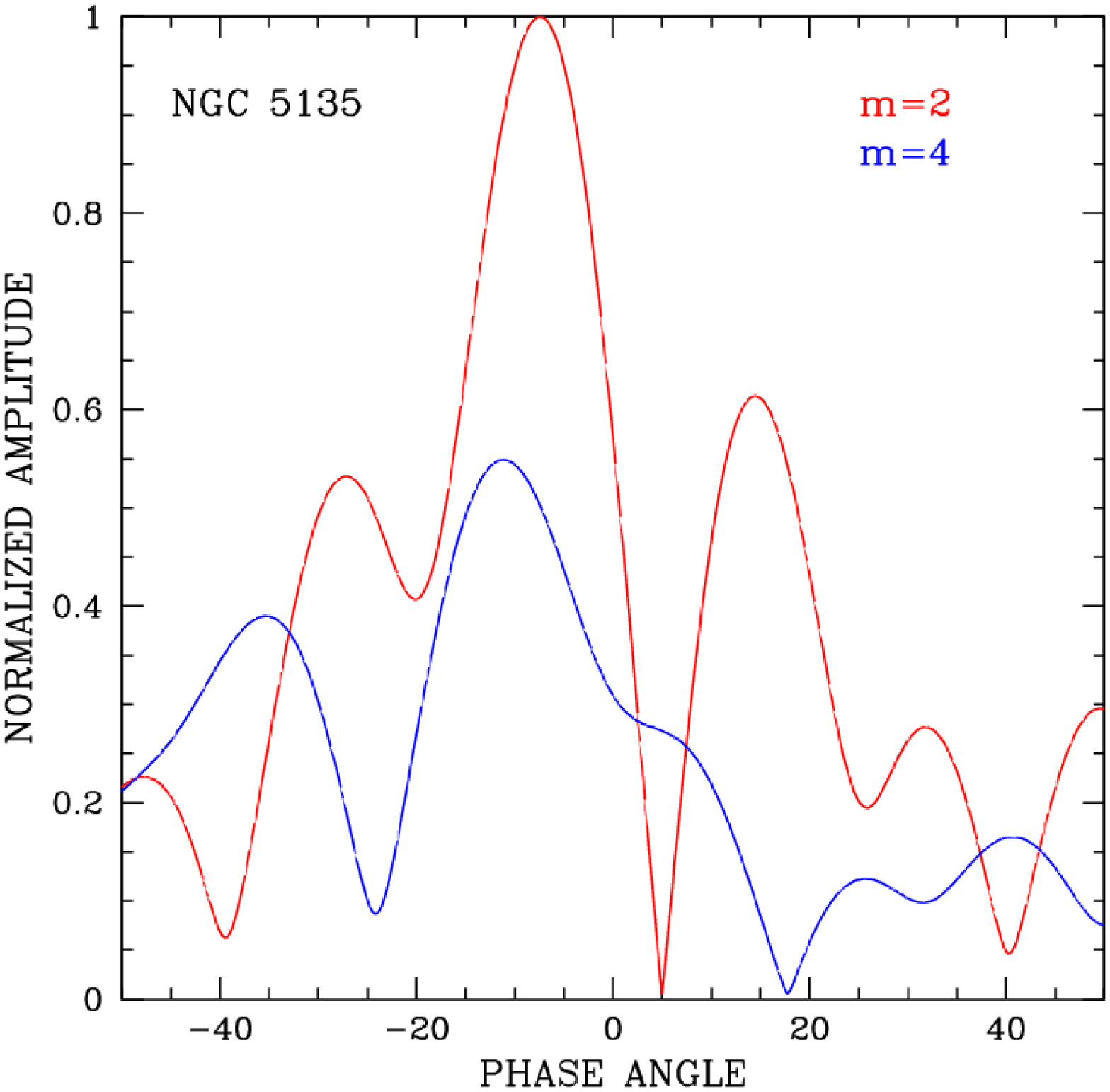}
\includegraphics{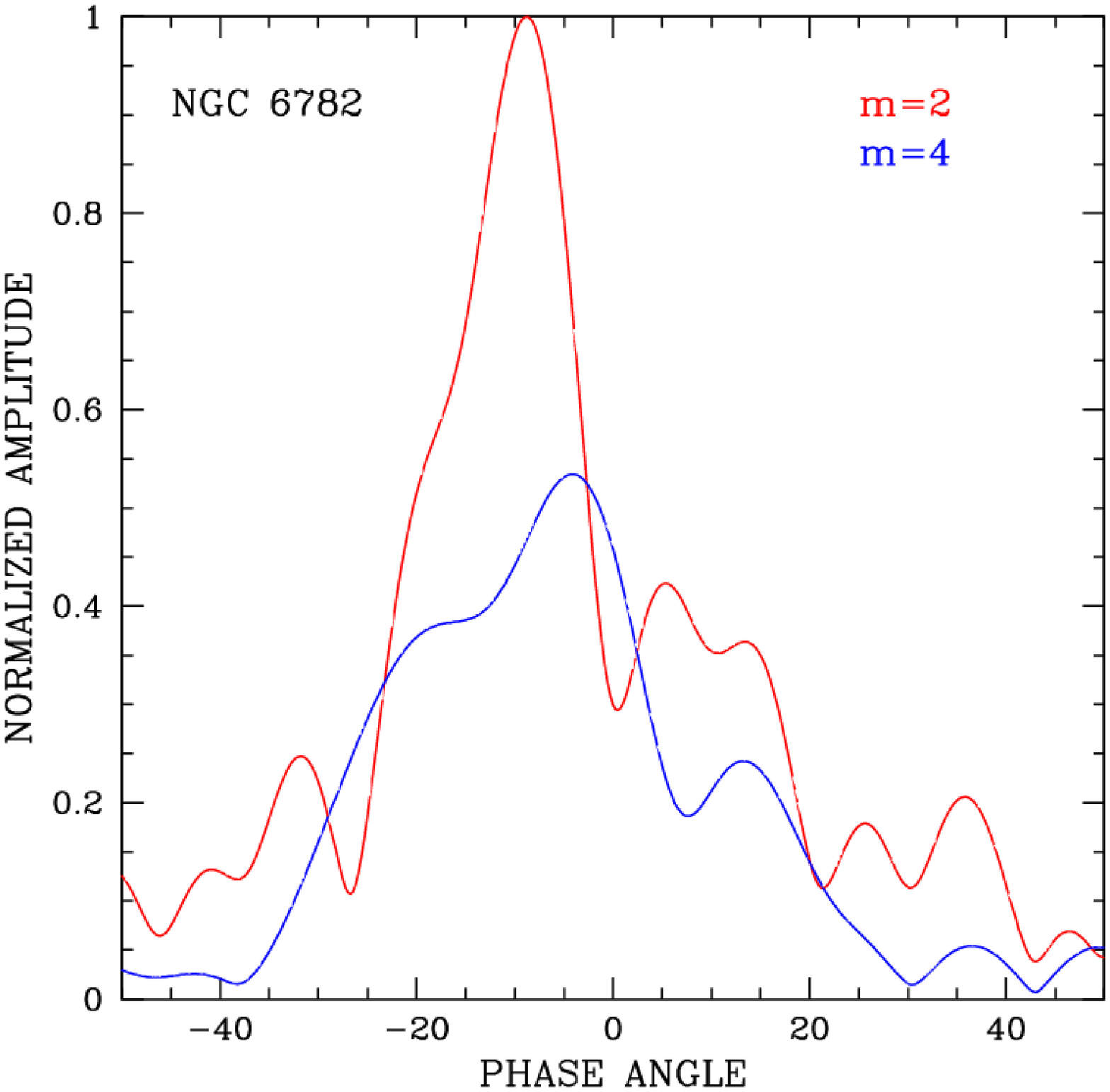}
\vspace*{8.5cm}
\caption{Plots of the relative amplitudes of the Fourier $m=2$ and $m=4$ 
components versus phase angle for NGC 5135 (left) and NGC 6782 (right). Pitch 
angle measurements of the much weaker $m=4$ mode are less reliable than those 
of the stronger $m=2$ mode.}
\label{landfig}
\end{figure*}


If the assertion is true that galaxies with low dark halo concentrations host black holes less massive
than average for a given halo mass and the $M_{BH}-P$ relationship holds, then we can 
make a prediction of the minimum mass of our fast-bar galaxies.  Figure 7 shows our $\mathcal{R}$
estimates versus the total galaxy mass implied from the $M_{BH}-P$ (Seigar et al. 2008) and $M_{BH}-M_{tot}$ (Bandara et al. 2009)
relationships.  Here $M_{tot}$ is the total gravitational mass of the galaxy within the radius where the density profile of the 
galaxy exceeds the critical density of the universe by a factor of 200. 
The predicted total mass of those galaxies with convincingly fast bars is expected to be a minimum (see Table 1).
Follow-up rotation curve measurements will test this prediction.

\begin{figure*}
\includegraphics{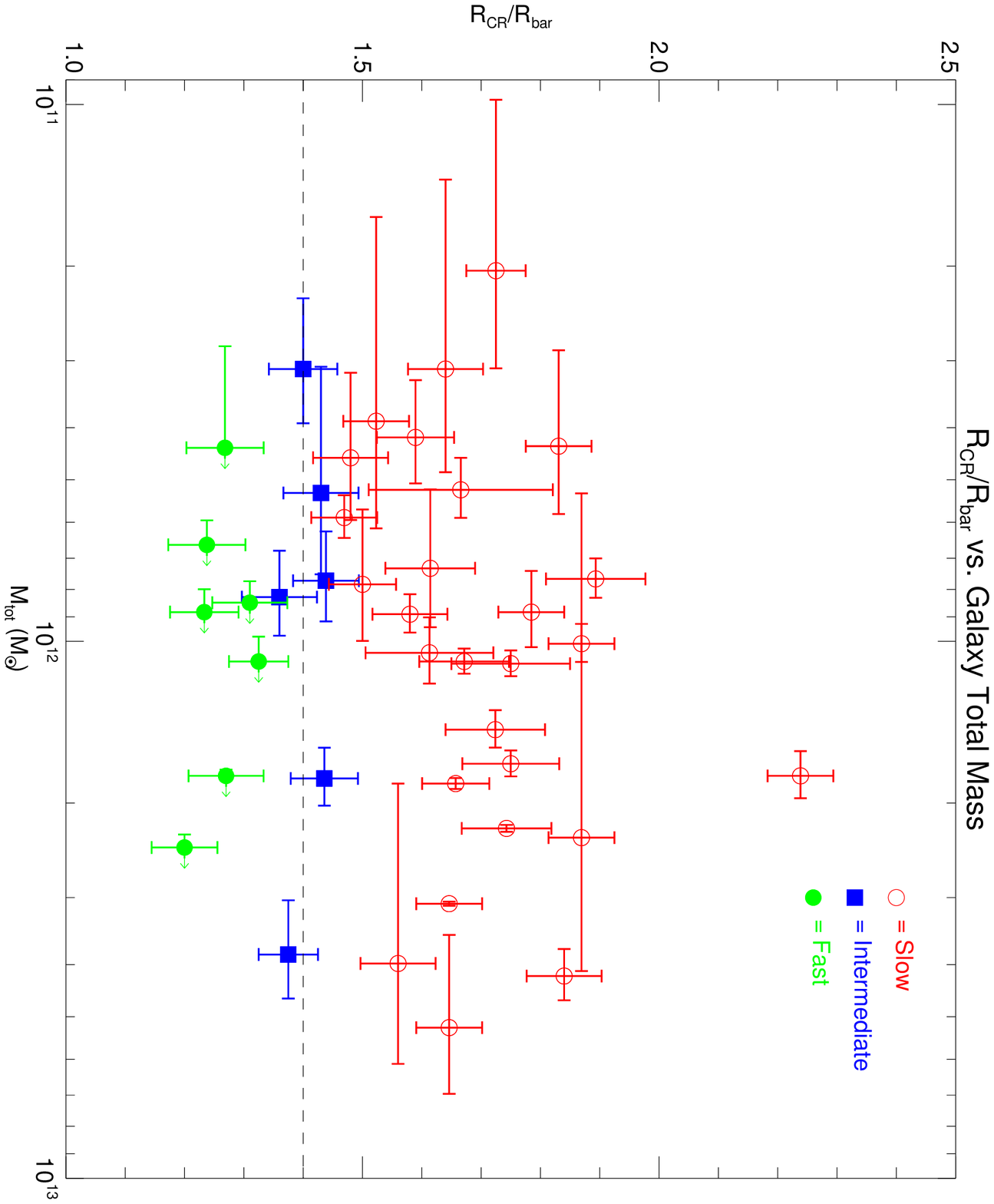}
\vspace*{8.5cm}
\caption{Plot of $\mathcal{R} = R_{CR}/R_{bar}$ versus total galaxy mass 
showing the distribution of total masses for galaxies.  The horizontal 
dashed line corresponds to 
$\mathcal{R} = 1.4$, or the boundary between the fast and slow bar domains}
\label{landfig}
\end{figure*} 

A plot of $P$ versus the de Vaucouleurs morphological type index ($T$; 
see Figure 8) shows a 
trend in agreement with Kennicutt (1981) and Seigar \& James 
(1998b).  
Galaxies identified as early-type have 
correspondingly small values of 
$P$ while late-types have larger values of $P$.  The trend becomes
a tighter correlation when the focus is shifted from the overall sample to only
those galaxies with clearly fast bars. A linear fit to the data of Roberts et 
al.\ (1975) concerning theoretical pitch angle versus Hubble type
is also overlayed (dot-dashed line).  The y-intercept of this fit 
was systematically lower than our data points, so it was multiplied by an ad hoc
factor of $\approx$ -0.46.  The slope of this theoretical fit is more 
similar to the slope of the linear fit of the galaxies with fast bars 
(dashed green line) than to
the slope of the linear fit of the entire sample (dotted line).

\begin{figure*}
\includegraphics{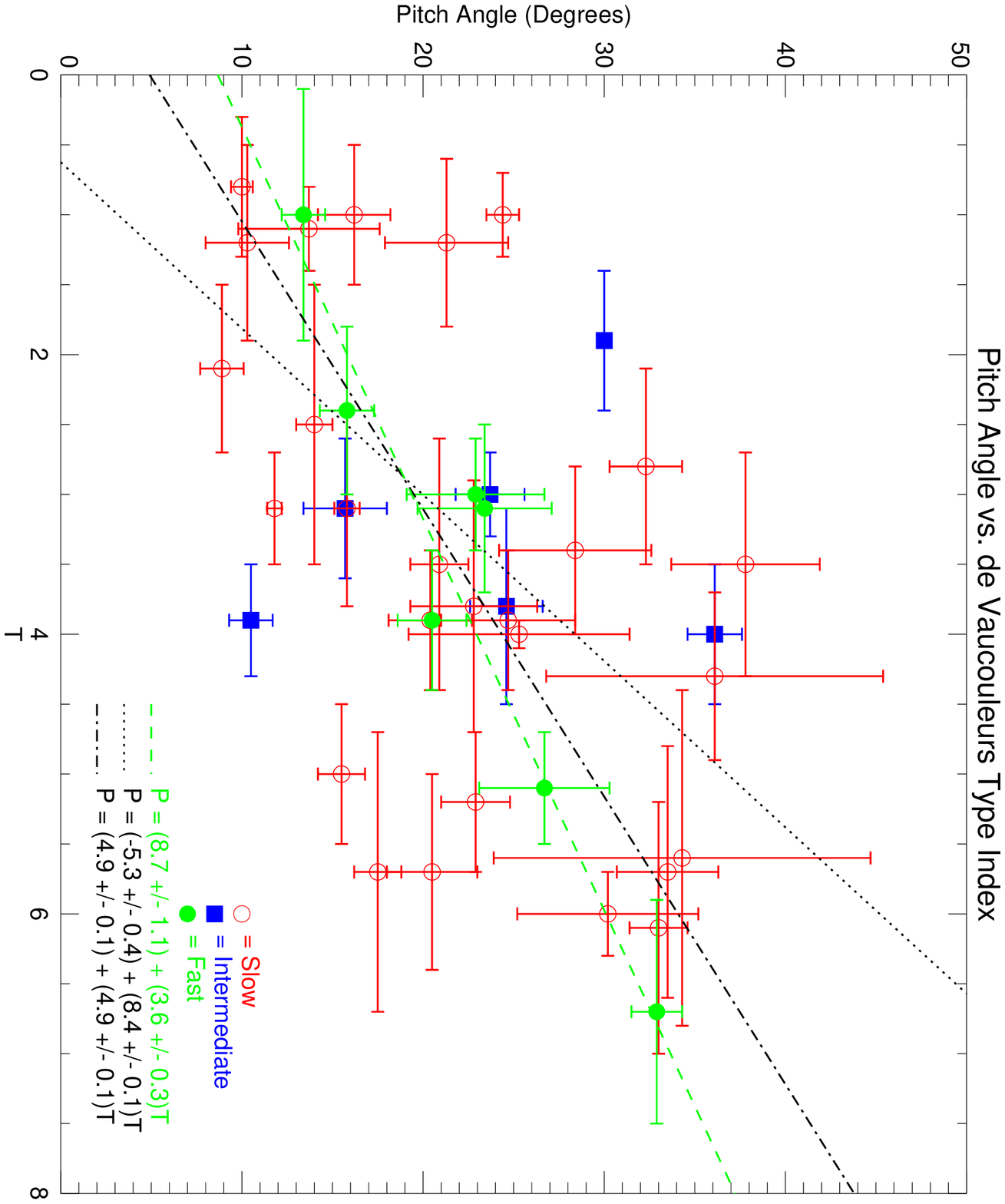}
\vspace*{8.5cm}
\caption{A plot of the measured spiral arm pitch angle versus de Vaucouleurs 
morphological 
type index taken from HyperLeda. The dashed green line is a fit through the 
points 
corresponding to galaxies with fast bars, the dotted line is a fit through all 
the 
points, and the dot-dashed line is the fit predicted by 
Roberts et al. (1975). The former two fits were obtained with the Interactive
Data Language (IDL) program
FITEXY. The latter, theoretical fit was systematically 
lower than the other two observationally derived fits. The y-intercept of the 
dot-dashed line was, therefore, scaled by a 
factor of $\approx$ -0.46 so that all three lines intersect at T $\approx$ 2.9.}
\label{landfig}
\end{figure*}

Das et al. (2008) have shown that a trend exists where bar strength ($Q_{g}$) 
is
anticorrelated with the central velocity dispersion ($\sigma_{c}$) of the bulge.
This result suggests that the growth of a central mass may be closely linked
to the evolution of the bar.  Additionally, Seigar et al. (2008) show that a 
good 
anticorrelation exists between $\sigma_{c}$ and $P$.  This is 
a result of $\sigma_{c}$ also being an indicator of $M_{BH}$ (Gebhardt et al.
2000; Ferrarese \& Merritt 2000).  These results imply that a correlation
between bar
strength and $P$ should be expected. In Figure 9 we plot $Q_{g}$ as a function
of $P$ and do not see this correlation in the overall sample.  However, when 
focusing only on the small sample of galaxies with clearly fast bars, $Q_{g}$ appears to 
increase with increasing $P$.  

\begin{figure*}
\includegraphics{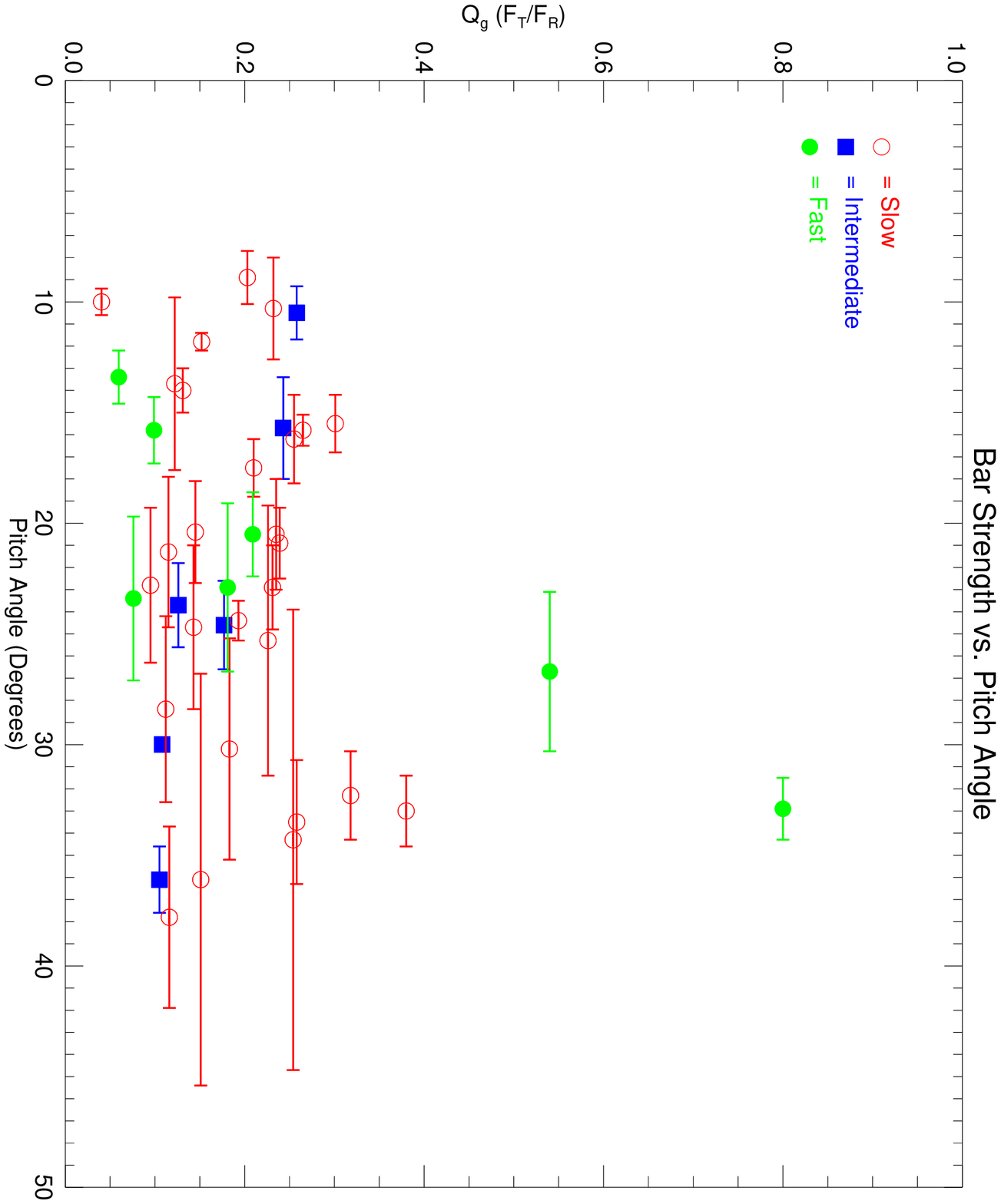}
\vspace*{8.5cm}
\caption{A plot of bar strength, $Q_{g}$, versus pitch angle. The anticorrelation of $Q_{g}$ versus
$\sigma_{c}$ shown in Das et al. (2008) is not evident in our overall sample when using the $\sigma_{c} - P$ relation 
(Seigar et al. 2008). However, the trend is seen when those galaxies with clearly fast bars are isolated.}
\label{landfig}
\end{figure*} 

An overall trend becomes more apparent when we plot
$Q_{B}$ as a function of $T$ for a larger sample of galaxies (see Figure 10). 
We have already shown that there is a loose 
correlation between $P$ and $T$ in general and perhaps a tighter correlation 
when only considering galaxies with fast bars.  Also, if $Q_{S}$
is small, then $Q_{g}$ approaches $Q_{B}$, effectively making $Q_{g}$ an upper
limit of $Q_{B}$.  
The larger overall sample displays a very weak correlation which agrees with the 
results of Laurikainen et al. (2007) and Seigar \& James (1998a), although the
latter authors determined the bar strength through an alternate method and 
plotted it as a function of bulge-to-disk ratio. When focusing once again on 
galaxies with clearly fast bars, it becomes more 
evident that a trend 
between bar strength and $T$, or $P$, exists for those galaxies.  This seems to suggest that galaxies with a smaller dark matter halo
contribution in the central regions follow a more predictable trend in bar strength and spiral arm morphology 
than those with more dominant dark matter contributions.

\begin{figure*}
\includegraphics{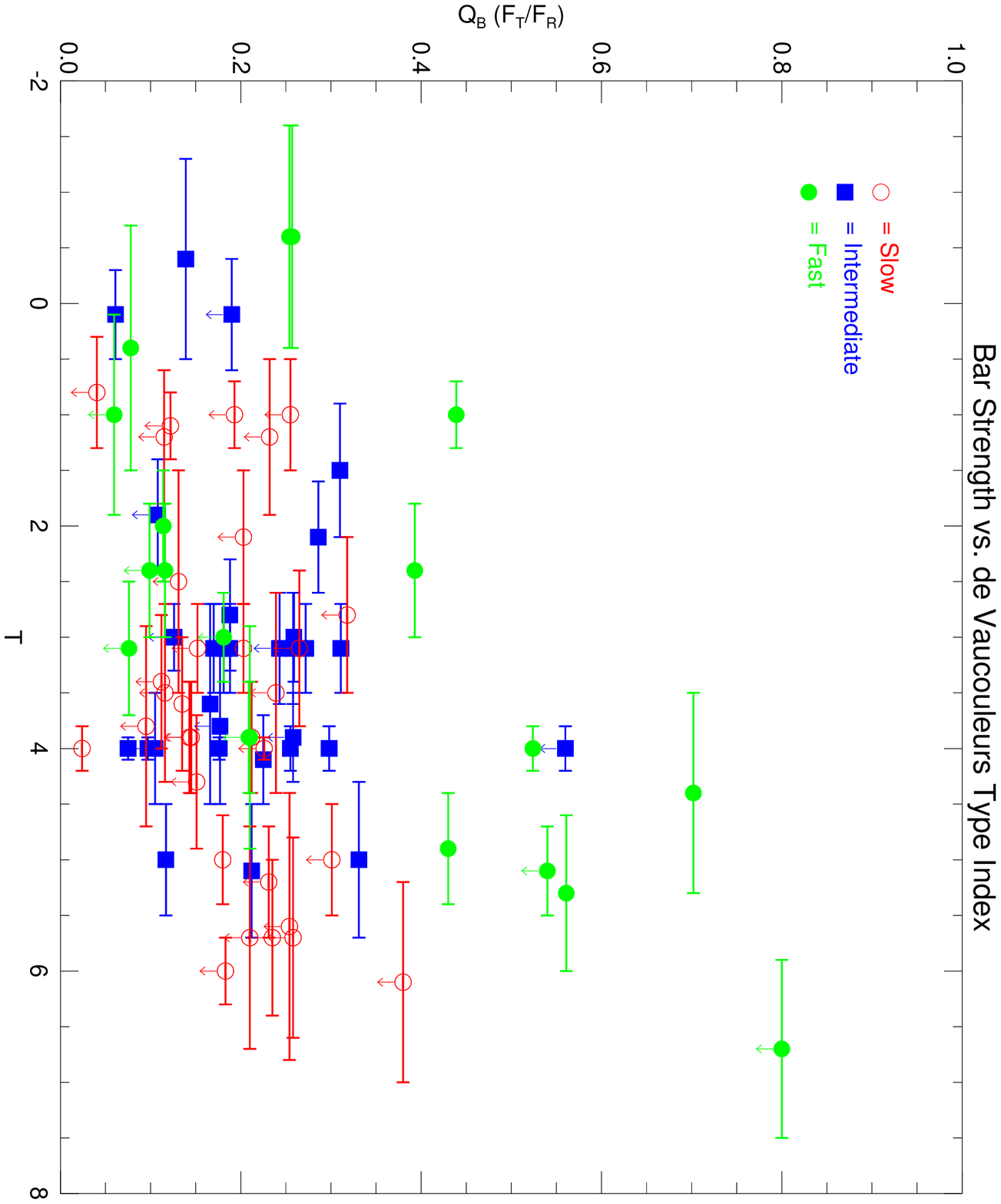}
\vspace*{8.5cm}
\caption{A plot of bar strength, $Q_{B}$, versus de Vaucouleurs 
morphological type index for a sample of four sets of galaxies including those
listed in: (1) Table 1 and their corresponding 
$Q_{g}$, which may be treated as an upper limit estimate of $Q_{B}$; (2) Table 
1 of Rautiainen et al. (2008; sans those included in Table 1 of this paper) 
and their corresponding $Q_{B}$
(Buta et al. 2005); (3) Table 2 of 
Rautiainen et al. (2008) where $Q_{B}$ was estimated by Buta et al. (2005); and
 (4) Treuthardt et al. (2008, 2009) where $Q_{g}$ was estimated by
the authors. The values of T were taken from HyperLeda for all of the galaxies.}
\label{landfig}
\end{figure*} 


Finally, Figure 11 displays $\mathcal{R}$ versus $T$ for our data as well as
that of Rautiainen et al.\ (2008). The general trend of $\mathcal{R}$ 
increasing with an increasing $T$, seen in Figure 6 of Rautiainen et al.\ 
(2008), is not as clear with the addition of our data. Instead, the few 
$\mathcal{R}$ values greater than $\sim$2.0, seen from $3 \la T \la 5$, appear 
to be anomalous rather than part of a general trend. The large range in $T$, or $P$, 
seen for slow, intermediate, and fast bar galaxies in this larger overall sample is indicative of the 
scatter around the empirical SMBH and halo mass relation of Bandara et al. (2009) predicted by Booth \& Schaye (2010).


\begin{figure*}
\includegraphics{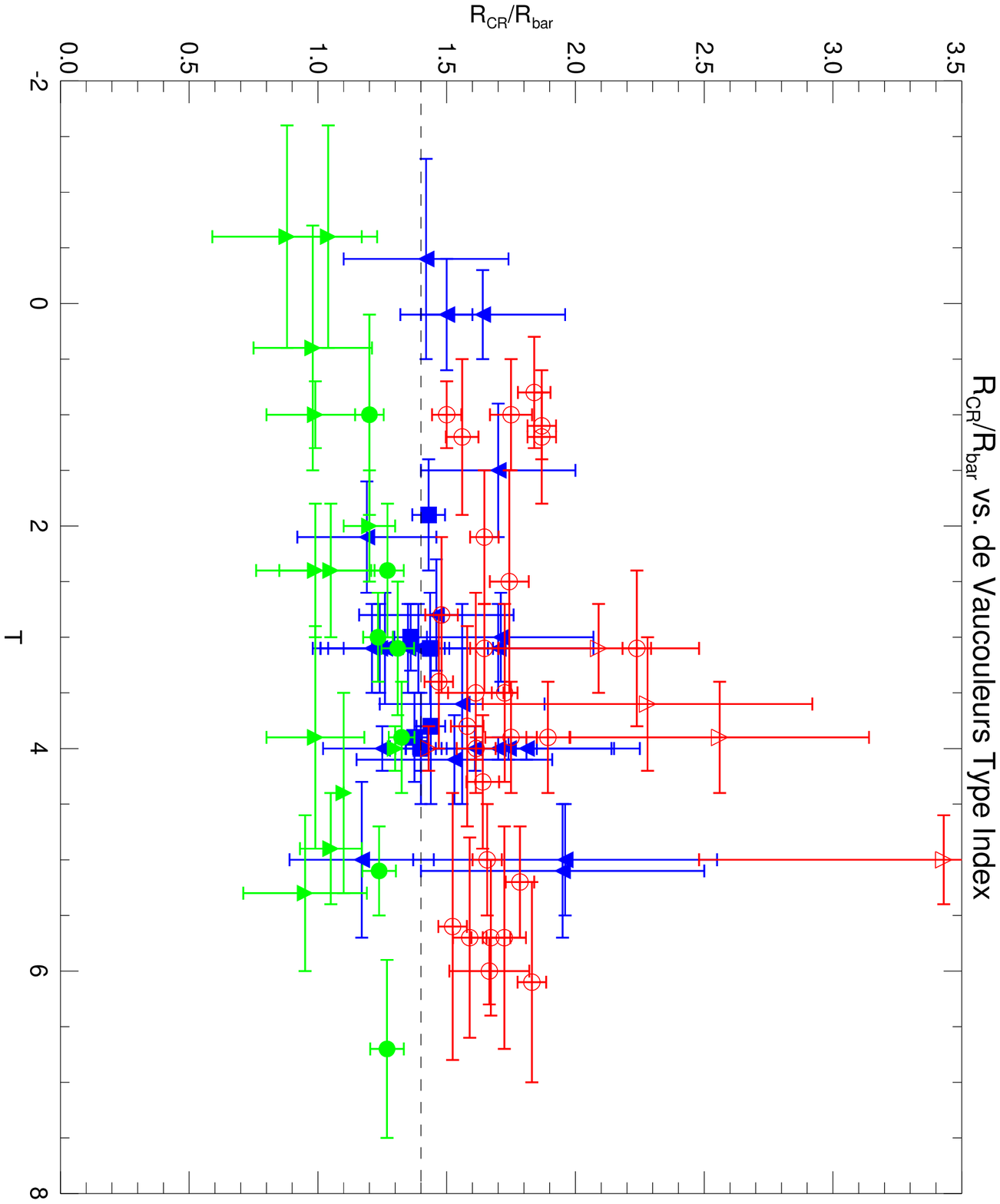}
\vspace*{8.5cm}
\caption{Plot of $\mathcal{R} = R_{CR}/R_{bar}$ versus $T$ with the horizontal 
dashed line corresponding to $\mathcal{R} = 1.4$, or the boundary between the 
fast and slow bar domains. The points are color coded as in Figure 10 where open circles and triangles correspond to galaxies with slow bars, filled circles and triangles to fast bars, and filled squares and inverted triangles to intermediate bars. The circles and squares correspond to data from this 
paper. The triangles correspond to $\mathcal{R}$ data from Rautiainen et 
al.\ (2008). The values of $T$ for all of the data points are 
taken from HyperLeda.}
\label{landfig}
\end{figure*}

\section{Conclusions}

We set out to estimate the SMBH mass of a sample of 40 barred spiral galaxies 
and determine which of these galaxies have low central dark halo densities. We 
did this by using a 2D FFT to measure $P$ and estimated $M_{BH}$ via the best
model fit of the $M_{BH}-P$ relationship (Seigar et al.\ 2008). We also 
created simulation models of these galaxies with morphologies resembling that 
seen in the CGS $B$-band images by primarily 
varying $\Omega_{p}$, a key dynamical component, or similarly $\mathcal{R}$. The
results of these morphological matches allowed us to discern those galaxies 
with convincingly \emph{fast} rotating bars and correspondingly low central 
density dark matter halos.

Our results indicate that a wide range of $M_{BH}$ exists in 
both fast and slow bar galaxies. This appears to support the theoretical work of Booth \& Schaye (2010)
where the authors claim that the scatter in the empirical relationship between SMBH and halo mass (Bandara et al. 2009)
is due to the dark matter halo concentration.
Additionally, we find that galaxies 
with low central dark halo densities appear to follow more predictable trends 
in $P$ versus $T$ and $Q_{B}$ versus $T$ than barred galaxies in 
general. The empirical $M_{BH}-M_{tot}$ relationship (Bandara 
et al. 2009) also provided us with an estimate of $M_{tot}$ for our galaxies.  
For a given halo mass, we expect the SMBHs in 
galaxies with fast bars to have a less-than-average mass due to the low 
central dark 
halo density and the resulting anemic growth predicted by Booth \& Schaye (2010). This means that the total 
luminous and dark mass of these galaxies is expected to be the minimum of what would be 
measured observationally. This claim can easily be tested by obtaining 
detailed kinematic data and this will be done in a forthcoming paper. 

It should be noted that the argument by Debattista and Sellwood 
(2000) correlating galaxies harboring fast bars with low central density
dark matter halos is theoretical and has not yet been tested observationally 
due to the inherent difficulty in doing so. Additionally, while dynamical 
friction from higher central density dark matter halos may not yet have slowed
recently formed fast bars, it is unlikely that any bars in our sample
are newly formed. We presume this because observational evidence suggests that 
bars are long-lived structures 
(Jogee et al. 2004; Sheth et al. 2003) and the galaxies in our sample are 
local and therefore old. We also acknowledge that bars may not slow due to 
dynamical friction alone. Models have shown that massive bulges can exchange
angular momentum with the bar and cause them to slow down (Athanassoula 2003).
If we assume that massive bulges reside in dark matter halos of high 
central concentration (e.g. Ho 2007), then angular momentum exchange with both
the bulge and halo would act to slow the bar. In this study, we are only 
interested in those galaxies with clearly \emph{fast} bars where any 
contribution to bar slow down by dynamical friction via a dark matter halo is 
minimal. 

The authors wish to thank B. Catinella and I. M\'{a}rquez for providing 
kinematical data for NGC 4050, NGC 3660, and NGC 5728. This work was supported 
by a grant through the Arkansas NASA EPSCoR program
and in part by the National Science Foundation under Grant CRI CNS-0855248, 
Grant EPS-0701890, Grant MRI CNS-0619069, and OISE-0729792. H. Salo 
acknowledges the support of the Academy of Finland. 
We also acknowledge the use of the HyperLeda database 
(http://leda.univ-lyon1.fr) and the NASA/IPAC Extragalactic Database (NED) 
which is operated by the Jet Propulsion Laboratory, California Institute of 
Technology, under contract with the National Aeronautics and Space 
Administration. The observational data presented in this paper were collected 
as part of the
Carnegie-Irvine Galaxy Survey (CGS; http://cgs.obs.carnegiescience.edu),
using facilities at Las Campanas Observatory, Carnegie Institution for
Science. The data were reduced independently from those
presented in Ho et al. (2011)

\label{lastpage}

\end{document}